\pdfoutput=1

\documentclass[11pt]{article}

\usepackage[final]{acl}

\usepackage{times}
\usepackage{latexsym}

\usepackage[T1]{fontenc}

\usepackage[utf8]{inputenc}

\usepackage{microtype}

\usepackage{inconsolata}

\usepackage{graphicx}

\usepackage{makecell}
\usepackage{subfigure}
\usepackage{xspace}
\usepackage{float}
\usepackage{tipa}
\usepackage{color}

\usepackage{arydshln}
\usepackage{enumitem}
\usepackage{multirow}
\usepackage{tcolorbox}
\usepackage{colortbl}
\usepackage{booktabs}

\usepackage[T1]{fontenc}
\usepackage{tabularx}
\usepackage{threeparttable}
\usepackage{pifont}
\usepackage[ruled,vlined]{algorithm2e}
\usepackage{algpseudocode}
\usepackage{setspace}
\usepackage{amsmath} 
\usepackage{multicol}

\newcommand{\ie}{{\em i.e.}}
\newcommand{\eg}{{\em e.g.}}
\newcommand{\cmark}{\ding{51}}%
\newcommand{\xmark}{\ding{55}}%
\newcommand{\methodname}{HalluHunter\xspace}

\title{Identifying the Achilles' Heel: An Iterative Method for Dynamically Uncovering Factual Errors in Large Language Models}

\author{Wenxuan Wang$^{1}$ \quad Yuk-Kit Chan$^{2}$\thanks{~~Denote Equal Contribution.}  \quad Zixuan Ling$^{2}$$^*$  \quad Juluan Shi$^2$$^*$ \quad Youliang Yuan$^{3}$  \\ \bf  Jen-tse Huang $^{4}$ \quad \bf Yifei Zhang$^{5}$ \quad \bf Wenxiang Jiao$^{6}$ \quad \bf Zhaopeng Tu$^{7}$ \quad \bf Michael R. Lyu$^2$ \\
$^1$Renmin University of China  \quad \quad $^2$The Chinese University of Hong Kong\\
$^3$The Chinese University of Hong Kong, Shenzhen \\
$^4$Johns Hopkins University \quad $^5$Nanyang Technological University \\
$^6$Xiaohongshu Inc. \quad $^7$Tencent Inc. \\  
$^1$\texttt{wangwenxuan@ruc.edu.cn} \quad \quad $^4$\texttt{jhuan236@jh.edu} \quad \quad  $^5$\texttt{wenxiangjiaonju@gmail.com} \\
}

\begin{document}
\maketitle
\begin{abstract}
Large Language Models (LLMs) like ChatGPT are foundational in various applications due to their extensive knowledge from pre-training and fine-tuning.
Despite this, they are prone to generating factual and commonsense errors, raising concerns in critical areas like healthcare, journalism, and education to mislead users.
Current methods for evaluating LLMs' veracity are limited by the need for extensive human labor, test data contamination, or limited scope, hindering efficient and effective exposure of errors. 
To address these challenges, we propose \methodname, a novel, fully automated framework for systematically uncovering factual inaccuracies in LLMs. \methodname employs a knowledge-graph-based approach, extracting fact triplets to generate diverse question types for single- and multi-hop reasoning using rule-based Natural Language Processing (NLP) techniques. Its iterative process starts with random triplet selection for question generation, followed by adaptive selection in subsequent iterations, targeting triplets where LLMs frequently err based on their performance analysis.
Our extensive tests on nine prominent LLMs reveal that \methodname can trigger factual errors in up to 55\% of tested questions.
Moreover, we demonstrate that \methodname’s test cases, particularly in adaptive selection, could further expose the weaknesses in benchmarking the factuality in LLMs meanwhile maintaining the coverage of questions. All code, data, and results are available at this link\footnote{https://github.com/Mysterchan/HalluHunter}.
\end{abstract}

\section{Introduction}
\label{sec-introduction}

Recent advancements in LLMs have propelled artificial intelligence to a notable milestone.
For example, ChatGPT has become one of the most prominent LLMs, demonstrating rapid adoption with 100 million monthly active users within two months of its launch, making it the fastest-growing software in history~\cite{chatgpt_fast}.
Moreover, LLMs have significantly impacted various applications, including machine translation ~\cite{jiao2023chatgpt}, grammatical error correction~\cite{Wu2023ChatGPTOG} and program synthesis~\cite{Gao2023ConstructingEI}.

\begin{table*}[t]
    \centering
    \caption{A comparison of \methodname to other factual evaluation works on the issues of high cost, data contamination, limited coverage and lack of effective expose of error}
    \label{table:related_works}
    \resizebox{1.0\textwidth}{!}{
    \begin{tabular}{l c c c c c c c }
    \toprule
    \bf \multirow{2}{*}{Dataset} & \bf Auto & \bf Dynamic &\bf Effective & \bf Question  & \bf Multi & \bf Cover Any & \bf LLMs \\
    & \bf Gen? & \bf Gen? &\bf Gen? & \bf Types  & \bf -Hop? & \bf Topics? &  \bf Tested?\\
    \midrule
    LAMA Probe~\cite{Petroni2019LanguageMA}& \xmark & \xmark & \xmark &1  & \xmark & \xmark & \xmark \\
    MLAMA~\cite{kassneretal2021MLAMA} & \xmark & \xmark & \xmark &1  & \xmark & \xmark & \xmark \\
    GrailQA~\cite{Gu2021GrailQA}  & \xmark  & \xmark &\xmark & 1  & \cmark & \xmark &  \xmark\\
    ParaRel  ~\cite{elazar2021ParaRel}& \xmark & \xmark & \xmark &1  & \xmark & \xmark  & \xmark\\
    SQuAD2.0~\cite{wang-etal-2021-generative} & \xmark &  \xmark   & \xmark & 1  & \cmark & \xmark & \xmark \\
    SimpleQuestions~\cite{wang-etal-2021-generative}  & \xmark  & \xmark &\xmark & 1   & \xmark & \xmark & \xmark\\
    KQA Pro~\cite{caoetal2022kqa} & \cmark  & \cmark &\xmark & 1   & \cmark & \xmark & \xmark\\
    PopQA~\cite{mallen2023PopQA} & \xmark  & \xmark &\xmark & 1   & \xmark & \xmark & \cmark\\
    PAQ~\cite{Lewis2021PAQ6M} & \cmark & \xmark  &\xmark & 1  & \xmark & \xmark & \xmark \\
    TruthfulQA~\cite{Lin2021TruthfulQAMH} & \xmark  & \xmark &\xmark & 1  & \xmark & \xmark &  \cmark\\
    SimpleQA~\cite{wei2024SimpleQa}  & \xmark  & \xmark &\xmark & 1  & \xmark & \xmark &  \cmark\\
    \newcite{Omar2023ChatGPTVT} & \xmark  & \xmark &\xmark & 2  & \xmark & \xmark &  \cmark\\
    Head-to-Tail~\cite{Sun2023HeadtoTailHK}  & \cmark & \cmark & \xmark &1  & \xmark & \xmark &  \cmark \\
    DyKnow~\cite{mousavi2024dyknow} & \cmark  & \xmark &\xmark & 1   & \xmark & \xmark & \cmark\\
    \midrule
    \bf Ours & \cmark & \cmark & \cmark & 3  & \cmark & \cmark & \cmark \\
    \bottomrule
    \end{tabular}
    }
\end{table*}

A significant barrier to the development of LLM-based intelligent applications, such as intelligence tutoring system, is their intrinsic proneness to errors, particularly in factual accuracy.
Prior studies, for instance, have shown that models like ChatGPT often produce plausible yet factually incorrect or nonsensical outputs, a phenomenon known as ``hallucinations''~\cite{Bang2023AMM}.
As these models advance and user trust in their outputs increases, such inaccuracies could lead to more serious consequences. This is especially problematic in sectors like journalism, academia, and healthcare where accuracy and reliability are paramount.
Therefore, identifying, analyzing, and mitigating these factual inaccuracies is essential to improve the safety and dependability of LLM-based intelligent software. 

While the progress on benchmarking the factuality in LLMs made by recent works is noteworthy, the current methods in evaluating factual accuracy have several shortcomings that require attention, as shown in Table~\ref{table:related_works}. These limitations hinder the ability to explore and address factual errors in LLMs:

\textbf{1. High Cost}: Existing benchmarks~\cite{Lin2021TruthfulQAMH, Talmor2019CommonsenseQAAQ, Laskar2023ASS} rely heavily on question formulation and human annotation, demanding significant effort. 
\textbf{2. Data Contamination}: LLM evaluation often suffers from data contamination due to the static nature of evaluation datasets, making the results unreliable. Unlike earlier models, LLMs use extensive internet-sourced corpora, potentially including publicly available evaluation data~\cite{Aiyappa2023CanWT, OpenAI2023GPT4TR}.
\textbf{3. Limited Coverage}: Prior research methods exhibit limitations in topic and question type, such as often focusing narrowly on specific relations like individuals and their birthplaces~\cite{Petroni2019LanguageMA, Kassner2021MultilingualLI}, or only using multiple choice questions, which have been shown to be a biased evaluation method~\cite{Li2024CanMQ}.
\textbf{4. Ineffective Error Exposure Mechanisms}: Current methods rely on single-round evaluations with static datasets, lacking effective mechanisms to generate questions that dynamically target areas where LLMs are prone to factual inaccuracies, limiting their ability to expose model weaknesses.


To address these limitations, we propose \methodname, a novel, fully automated framework designed to comprehensively evaluate and uncover factual inaccuracies in LLMs. \methodname leverages a knowledge-graph-based and rule-based NLP approach, by selecting facts triplets to generate diverse question types, including Yes/No, Multiple-Choice (MC), and WH questions and supporting both single-hop and multiple hops reasoning. By eliminating reliance on manual annotation and static datasets, \methodname reduces costs and mitigates data contamination risks with questions generated dynamically at test-time. Its flexible design allows customization across various domains and knowledge bases, ensuring broad topic coverage.

\begin{figure*}[t]
    \centering
    \includegraphics[width=\linewidth]{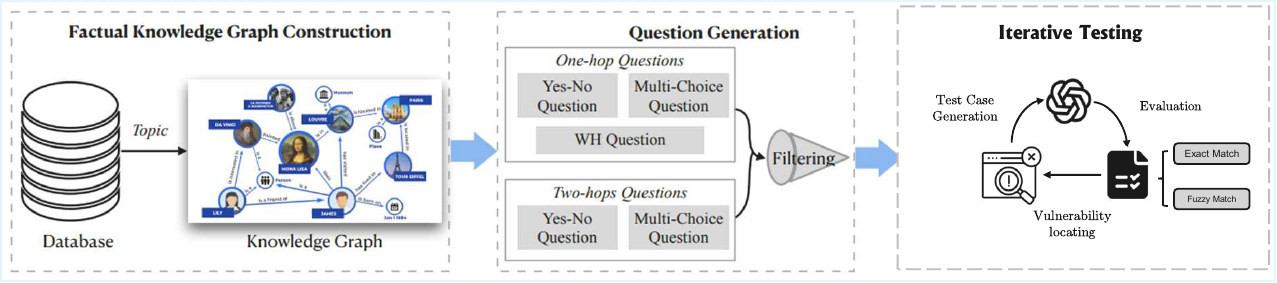}
    \caption{An illustration of the framework of \methodname.}
    \label{fig:architecture}
\end{figure*}

A core innovation of \methodname is its iterative and adaptive test case generation algorithm (Algorithm~\ref{alg:factprobe}), which analyzes LLM performance from prior iterations to dynamically select fact triplets entities structurally similar to those in incorrect responses and relations with low accuracy. This adaptive approach enhances the framework’s ability to generate targeted questions that expose LLM inaccuracies across diverse domains, iteratively increasing test difficulty to probe deeper vulnerabilities. The architecture is illustrated in Figure~\ref{fig:architecture}.


To comprehensively assess the effectiveness of \methodname, we evaluate it on nine widely deployed LLMs: GPT-3.5-Turbo, GPT-4-Turbo, GPT-4o, DeepSeek-v3, Claude-Sonnet-4, Claude-3.5-Haiku, Gemini-2.0-Flash, Qwen-3, and Qwen-3-Reasoning. Our evaluation shows that \methodname successfully uncovers factual inaccuracies in these models. The key contributions of this work are:
\begin{itemize}
    \item We design and implement \methodname, a novel, fully automated framework that exposes factual inaccuracies in LLMs by generating diverse question types using knowledge graphs, eliminating reliance on human annotation and mitigating data contamination risks.
    
    \item We develop an iterative and adaptive algorithm within \methodname that dynamically generates questions by selecting fact triplets based on prior LLM performance, targeting entity- and relation-specific weaknesses to show deeper vulnerabilities.
    
    \item We conduct an comprehensive evaluation of \methodname using random and iterative question generation across nine state-of-the-art LLMs, demonstrating their ability to identify and expose significant factual errors.
\end{itemize}

\section{Approach and Implementation}
\label{sec-method}

In this section, we present \methodname, a novel framework designed to identify factual errors in LLMs. The architecture comprises four stages:
\begin{enumerate}[leftmargin=*]
    \item \textit{Knowledge Graph Construction}: Building a structured knowledge graph (KG) from fact triplets sourced from an external database.
    \item \textit{Random Question Generation}: Producing diverse one-hop and multi-hop questions by randomly selecting triplets from the KG.
    \item \textit{Answer Assessment}: Querying target LLMs and identifying potential factual errors through matching algorithms.
    \item \textit{Iterative and Adaptive Question Generation}: Employing an iterative and adaptive algorithm that dynamically updates the question generation strategy based on evaluation results from prior rounds, effectively selecting triplets to target LLM vulnerabilities and enhance the efficiency of exposing factual inaccuracies in subsequent iterations.
\end{enumerate}


\subsection{Knowledge Graph Construction}
\label{sec:construct_kg}

The first step in \methodname is the construction of a structured factual knowledge graph (KG), which serves as the foundation for generating evaluation questions. The primary contribution of \methodname lies in its flexible and robust framework design, which can seamlessly integrate with various knowledge bases to extract fact triplets. For this study, we demonstrate the framework's capability using Wikidata, a widely accessible knowledge base with over 100 million items~\footnote{\url{https://www.wikidata.org/wiki/Wikidata:Main_Page}}. 

\methodname enables users to focus on specific topics by defining criteria, such as ``occupation: emperor.'' These criteria are translated into SPARQL queries~\footnote{\url{https://www.wikidata.org/wiki/Wikidata:SPARQL_query_service}} to retrieve relevant triplets from the chosen knowledge base. The selected fact triplets are represented as (SUBJECT, relation, OBJECT), for example, (USA, capital, Washington D.C.), indicating that Washington D.C. is the capital of the USA. 

After retrieving the triplets, a directed graph is constructed by \methodname, denoted as $G = (V, E)$, where the vertex set $V$ comprises SUBJECT and OBJECT entities, and the edges in $E$ represent relations pointing from the SUBJECT vertex to the OBJECT vertex.

\subsection{Question Generation}
\label{sec:question_generation}

\methodname employs a systematic rule-based approach to generate questions from the KG, supporting Yes-No, MC, and WH question types, including single-hop and multi-hop complexities. Implementation details and templates are in Appendix~\ref{app:rule_based_explaination}.
\subsubsection{One-Hop Questions Generation}

For each randomly selected triplet, \methodname creates targeted questions, supporting all major English question types, as shown in Table~\ref{tab:question_example} and Figure~\ref{fig:rule-based}.

\begin{table*}[t]
\centering
\caption{Examples of generated questions. The first column shows single-hop questions while the second column shows multi-hop ones.}
\resizebox{1.0\textwidth}{!}{
\begin{tabular}{l l l l}
\toprule
\bf Tuple & \bf Type   & \bf Question  & \bf  Answer\\
\midrule
\multirow{4}{*}{(Napoleon, native language, Corsican) } &  Yes-No  & Is Corsican the native language of Napoleon?  & Yes\\
\cmidrule(lr){2-4}
 &  MC  & \makecell[l]{What is the native language of Napoleon? \\A. Latin  B. Chinese  C. Corsican  D. Marathi } & C \\
 \cmidrule(lr){2-4}
& WH  & What is the native language of Napoleon?  & Corsican\\
\midrule
(Michelle Obama, spouse,  & Yes-No & Was Michelle Obama's spouse educated at Harvard University? & Yes \\
\cmidrule(lr){2-4}
educated at, Harvard University)&  MC  &  \makecell[l]{Where was Michelle Obama's spouse educated at? \\ A. Harvard University  B. UCLA  C. Stanford University  D. MIT } & A\\
\bottomrule
\end{tabular}}
\label{tab:question_example}
\end{table*}

\textbf{Yes-No Questions}: Using a triplet (Subject, Relation, Object), \methodname conducts Part-of-Speech (PoS) analysis on the relation to select an appropriate auxiliary verb (e.g., ``is'' for nouns, ``does'' for verbs), yielding questions like ``Is Washington D.C. the capital of the USA?'' from (USA, capital, Washington D.C.). For passive verbs, the structure adjusts for clarity.

To ensure balanced testing, \methodname generates an equal number of ``No''-answer questions by selecting an incorrect but relation-consistent Object, such as ``Is London the capital of the USA?''

\textbf{Multiple-Choice Questions}: \methodname uses Named Entity Recognition (NER) to select interrogative pronouns (e.g., ``What,'' ``Who,'' ``Where'') for SUBJECT or OBJECT queries. For SUBJECT queries, PoS analysis determines the auxiliary verb, forming questions like ``Which country's capital is Washington D.C.?'' For OBJECT queries, a similar approach applies.

For each triplet, \methodname generates four options: one correct answer and three distractors sourced from other knowledge graph edges with the same relation. For example, for (Donald Trump, child, Ivanka Trump), \methodname formulates ``Who is a child of Donald Trump?'' and selects distractors like Malia Obama (child of Barack Obama), Chelsea Clinton (child of Bill Clinton), and Jennifer Gates (child of Bill Gates), randomly assigned as options A, B, C, and D, to test the LLM’s ability to identify correct relationships.

\textbf{WH Questions:} \methodname has stricter requirements for fact triplets, to ensure that the questions have a unique answer in Wh questions.
For instance, instead of generating the question ``What is the city of China?'' for the fact triplet (China, city, Shanghai), it is more appropriate to generate the question ``What is the capital of China?'' based on the fact triplet (China, capital, Beijing).

To achieve the above requirement, \methodname only selects triplets with a single outgoing edge for the relation.
For example, for (China, city, Shanghai), when considering the source entity ``China,'' there are multiple out-edges labeled as ``city'' pointing to different cities. While for (China, capital, Beijing), there will only be one out-edge of ``China'' labeled as ``capital'' pointing to ``Beijing.'' 
By guaranteeing the uniqueness of the answer for the generated question, \methodname limits the variation in correct answers, making the final verification process much more straightforward.

\subsubsection{Multi-Hop Questions Generation}
\label{sec-multi-hops}

\methodname generates multi-hop questions that require chained reasoning, such as ``Where was Michelle Obama’s spouse educated at?'' from Table~\ref{tab:question_example}. These questions demand multiple steps, like identifying Michelle Obama’s spouse as Barack Obama and then determining his education at Harvard University.

For multi-hop questions, \methodname uses extended triplets (SUBJECT, relation-list, OBJECT), e.g., (Michelle Obama, \{spouse, educated at\}, Harvard University). It concatenates the SUBJECT with the first relation (e.g., ``Michelle Obama’s spouse'') and applies PoS analysis to the final relation to select the auxiliary verb, producing questions like ``Where was Michelle Obama’s spouse educated at?'' using the same generation process as one-hop questions.'

\subsection{Answer Assessment}
\label{sec:assess}

\subsubsection{LLM Responses Collection}

Once \methodname has generated a significant number of questions, we can utilize them as test cases to query LLMs. Detail Prompts are provided in Appendix~\ref{sec:prompt_detail}.

\subsubsection{LLM Errors Identification}
\label{sec-error-indentify}

After collecting LLM responses, \methodname evaluates performance to identify factual errors. \textbf{For Yes-No and MC questions}: accuracy is assessed via exact matching against ground-truth answers. \textbf{For WH questions}: to account for entity variations (e.g., ``Great Britain'' vs. ``United Kingdom''), \methodname employs five similarity metrics to determine response correctness:

\begin{itemize}[leftmargin=*]
    \item \textbf{Levenshtein distance}: It is a string metric that quantifies the minimum number of single-character edits required to transform one word into another \cite{Po2020SimilarityBI}. 
    \item \textbf{N-grams similarity}: It measures the similarity of two sequences by comparing the overlapping ratio of sub-sequences they contain~\cite{Papineni2002BleuAM}. We used N=1.
    \item \textbf{Word embedding similarity}: It measures the semantic similarity between words represented as dense vector embeddings in a high-dimensional space, adopted in \cite{Chen2021TestingYQ}.
    \item \textbf{Sentence transformer similarity}: It utilizes the sentence transformer model\footnote{https://github.com/UKPLab/sentence-transformers} to represent the whole sentences in a vector form. 
    \item \textbf{ChatGPT}: \methodname directly asks ChatGPT (gpt-3.5-turbo-0613) whether the LLM response is equivalent to the question answer.
\end{itemize}

The questions that can not be answered correctly by the LLMs will be collected as suspicious errors for further human analysis.

\subsection{Iterative and Adaptive Question Generation}

\setstretch{0.9} 
\begin{algorithm}[h]
\small 
\caption{Iterative and Adaptive Test Case Generation}
\label{alg:factprobe}
\SetKwInOut{Input}{Input}
\SetKwInOut{Output}{Output}

\Input{Knowledge graph $G = (V, E)$, question list $Q^{(l)}$, LLM response list $A^{(l)}$, triplet set $T^{(l)}$, Relation-accuracy $R^{(l)}$, embedding model $\mathcal{M}$, top-$k$ parameter $k$, explore constant $e$,  low-accuracy constant $a$}
\Output{New question list $Q^{(l+1)}$, Triplet set $T^{(l+1)}$, Relation-accuracy $R^{(l+1)}$}
Initialize new question list\;
$Q^{(l+1)} \gets \emptyset$\;
Remove used triplets to prevent duplication\;
$T^{(l+1)} \gets T^{(l)} \setminus \{t_i \mid q_i \in Q^{(l)}\}$ \;
$\text{Shuffle}(T^{(l+1)})$\;
$R^{(l+1)} = R^{(l)}$

\For{each $(q_i, a_i) \in (Q^{(l)}, A^{(l)})$ \text{with triplet} $t_i = (s_i, r_i, o_i)$}{
    $c_i \gets \text{Evaluator}(q_i, a_i, t_i)$ \Comment{$c_i$: True if response is correct, False otherwise} \;
    Update $R^{(l+1)}(r_i)$ \;
}

\For{each $(q_i, a_i) \in (Q^{(l)}, A^{(l)})$ with triplet $t_i = (s_i, r_i, o_i)$}{
    $e_i \gets \text{random}()$ \Comment{value between [0,1]} \;
    \uIf{$e_i < e$} 
    {
        Explore the triplets with low accuracy relation \;
        $T' \gets \{t' \in T^{(l+1)} \mid R^{(l+1)}(t') < a$\}\;
    }
    \uElse {
        \uIf{not $c_i$}{
            $C \gets \text{TopKSimilar}(s_i, k, \mathcal{M})$\;
            $T' \gets \{t' \in T^{(l+1)} \mid t'[0] \in C\}$\;
        }
        \uElse {
            $T' \gets T^{(l+1)}$ \;
        }
    }
    \For{$(s', r', o') \in T'$}{
        Generate question $q'$ for $(s', r', o')$ with $q_i.\text{type}$\;
        \If{$q' \neq \text{Null}$}{
            $Q^{(l+1)} \gets Q^{(l+1)} \cup \{q'\}$\;
            \textbf{break}\;
        }
    }
}
\Return $Q^{(l+1)}, T^{(l+1)}$ \;
\end{algorithm}

\setstretch{1}

The iterative and adaptive question generation stage is a cornerstone of the \methodname framework, designed to expose factual inaccuracies in LLMs by dynamically generating targeted questions based on prior iteration results. This stage integrates the question generation and answer assessment stages (Sections~\ref{sec:question_generation} and \ref{sec:assess}) to iteratively target the LLMs weaknesses. Formally, given a knowledge graph $G = (V, E)$, where $V$ represents entities and $T$ contains triplets $t = (s, r, o)$ with $s, o \in V$ as subject and object respectively and $r$ as a relation. This algorithm processes the question list \( Q^{(l)} \), LLM response list \( A^{(l)} \), triplet set \( T^{(l)} \), and relation-accuracy map \( R^{(l)} \) at iteration \( l \) to produce a new question list \( Q^{(l+1)} \), updated triplet set \( T^{(l+1)} \), and updated relation-accuracy map \( R^{(l+1)} \). 

The algorithm’s design is motivated by the hypothesis that LLM factual inaccuracies arise from unfamiliarity with specific entities or relations, treated as knowledge points in the knowledge graph. For example, if an LLM fails to answer a question based on the triplet (hydrogen, atomic mass, 1.008), it may also struggle with similar questions in chemistry, such as "What is the atomic mass for oxygen?" If it fails on a question from the triplet (13, prime factor, 91), it may lack understanding of the relational concept of prime factors. To address this, Algorithm~\ref{alg:factprobe} evaluates question-response pairs \( (q_i, a_i) \in (Q^{(l)}, A^{(l)}) \) with corresponding triplets \( t_i = (s_i, r_i, o_i) \), computing a boolean correctness indicator \( c_i\) (\text{True}  for correct responses and \( \text{False} \) otherwise) as described in Section~\ref{sec:assess}. It updates the relation-accuracy map \( R^{(l+1)} \) to track the running average performance of the tested model on different relations. The triplet set \( T^{(l+1)} \) is shuffled and depleted of used triplets to ensure diversity and avoid repetition. During each generation, with a fixed probability \( e \), the exploration constant and \(a\), defined as low accuracy, the algorithm selects triplets with low-accuracy relations (\( R^{(l+1)}(r_i) < a \)) with higher priority to exploit known weaknesses. 

If the previous response is incorrect (\( c_i = \text{False} \)), a knowledge graph embedding model \( \mathcal{M} \), trained using QuatE~\cite{zhang2019Quate} within the PyKEEN framework~\cite{ali2021pykeen}, will be used to identify entities similar to \( s_i \) in the embedding space to generate challenging questions that probe related knowledge points. New questions \( q' \) for triplets \( (s', r', o') \) preserve the type of \( q_i \). 
If the previous response is correct (\( c_i = \text{True} \)), a random valid triplet will be picked to generate a new question.
By balancing exploitation of weak relations and subject entity, the algorithm effectively exposes the LLM weaknesses in various domains.

\label{sec-approach}

\section{Evaluation}

\label{sec-experiment}

\subsection{Experimental Setup}

\paragraph{Software and Models Under Test}
To assess the effectiveness of the \methodname, we employ it to evaluate nine widely-utilized LLMs models: gpt-3.5-turbo-0125, gpt-4-turbo-2024-04-09, gpt-4o-2024-11-20, deepseek-v3-0324, claude-sonnet-4-20250514, claude-3-5-haiku-20241022, gemini-2.0-flash,  qwen3-32b and qwen3-32b-reasoning. All are with the default temperature.

We focus on a closed-book evaluation setting, where models answer without retrieval augmentation or external grounding. This setting is important in practice because many real-world interactions with LLMs remain weakly grounded, and even in grounded systems, parametric knowledge can still influence answer selection and reasoning behavior ~\cite{mallen2023PopQA, tan2025dynamicparametricretrievalaugmented}. Therefore, evaluating parametric knowledge remains a meaningful and complementary way to diagnose factual weaknesses in LLMs.

\paragraph{Test Cases Generation}
To comprehensively evaluate LLMs' performance, we conduct experiments by generating questions from three big domains: Humanity, Social Science and Science, technology, engineering, and mathematics (STEM). We use \methodname to generate 1000 questions for each question type within each domain. The iterative algorithm will generate 1000 new questions for each loop to support a same scale comparison.

\paragraph{Evaluation Metrics}
To rigorously assess performance of LLMs and our novel algorithm, we employ two primary metrics tailored to the objectives of \methodname:

\begin{itemize}[leftmargin=*]
    \item \textbf{Accuracy}: We measure the factual accuracy of each LLM by the percentage of correct answers. For Yes-No and MC questions, accuracy is determined through exact matching, For WH questions, accuracy is evaluated using different metrics as detailed in Section~\ref{sec:assess}.
    \item \textbf{Weighted Coverage}: Our goal is to ensure that our method maintains a balanced coverage of the entire knowledge graph, avoiding a focus on rare, unconventional facts that are difficult to reach. We aim to give more weight to entities that are more easily reachable—those that have a larger number of neighbors. 
    Inspired from Group Degree Centrality~\cite{group_degree_centrality}, a metric commonly used to evaluate the importance or centrality of groups of nodes in networks, we calculate the Group Degree Centrality of the entity set that is selected to form the final question set as its weighted coverage. 
    Let \(G=(V,E)\) be a graph and \(S\subseteq V\) the selected (visited) set; define the open neighborhood
    \(N(S)=\{\,v\in V\setminus S:\exists\,u\in S\text{ with }(u,v)\in E\,\}\).
    The (unnormalized) group degree centrality is \(C_{\deg}(S)=|N(S)|\), and we report its normalized form
    \(\widehat{C}_{\deg}(S)=\dfrac{|N(S)|}{|V|-|S|}\in[0,1]\).
    
\end{itemize}



\begin{figure}[t]
    \centering
    \includegraphics[width=\linewidth]{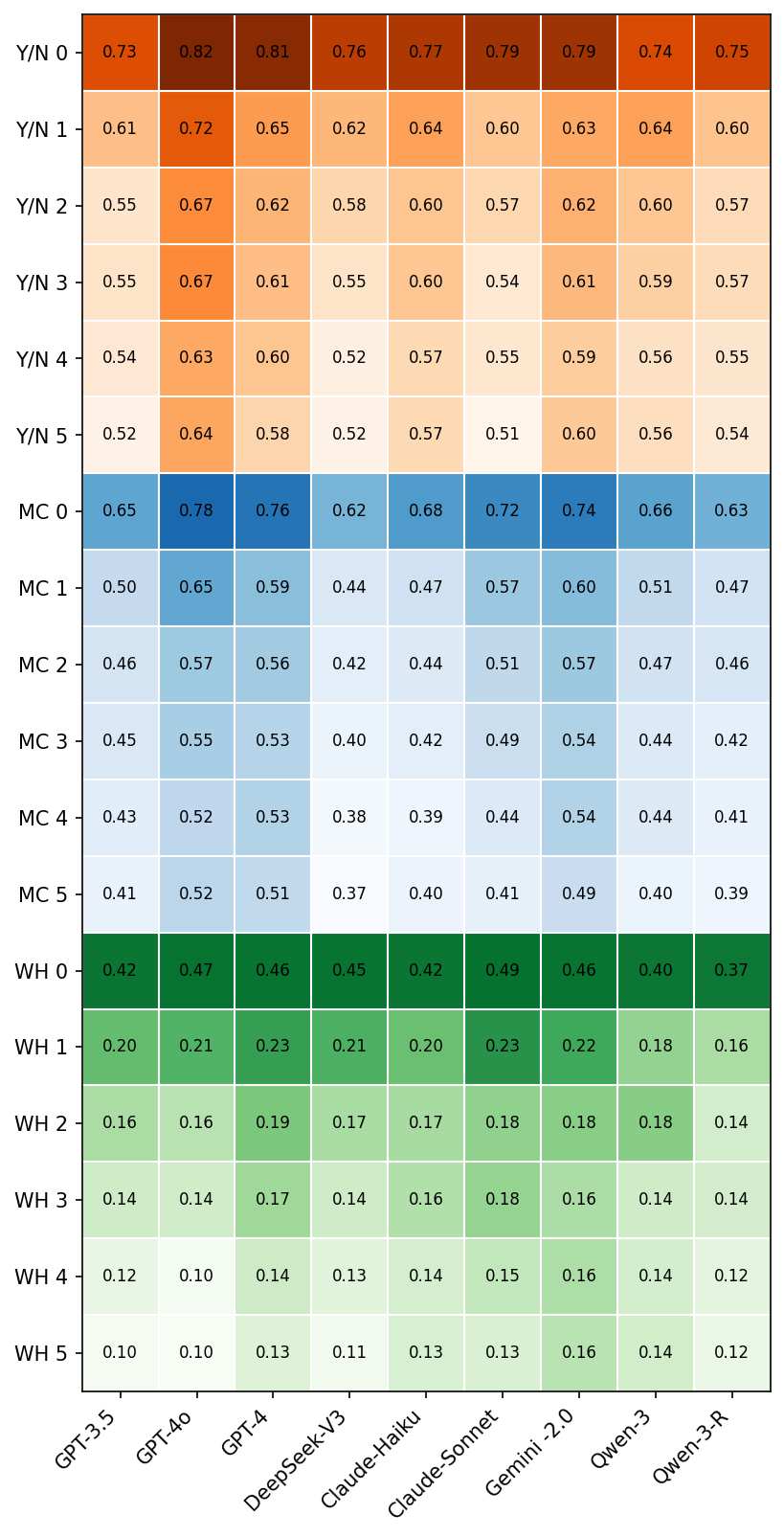}
    \caption{Heat map of LLMs' average accuracy across sequential trials in 1-Hop questions (darker shades indicate higher accuracy)}
    \label{fig:heatmap_grouped}
\end{figure}

\subsection{Baseline Performance Analysis}

This section evaluates the performance of LLMs under the initial random question generation phase of \methodname, referred to as Trial 0, to establish a baseline for factual accuracy across diverse question types and domains. Detailed numerical results are provided in Appendix~\ref{sec:result_detail}.

\textbf{\methodname reveals diverse factual errors in LLMs.} By generating and evaluating responses to random question sets, \methodname effectively identifies factual inaccuracies across different LLMs. As shown in Trial 0 of Figure~\ref{fig:heatmap_grouped}, GPT-4o achieves the highest accuracy, followed by GPT-4. Notably, WH questions pose the greatest challenge, with an average accuracy of 37.4\% across all LLMs, indicating their difficulty in precise factual recall.

\begin{figure}[t]
    \centering
    \includegraphics[width=\linewidth]{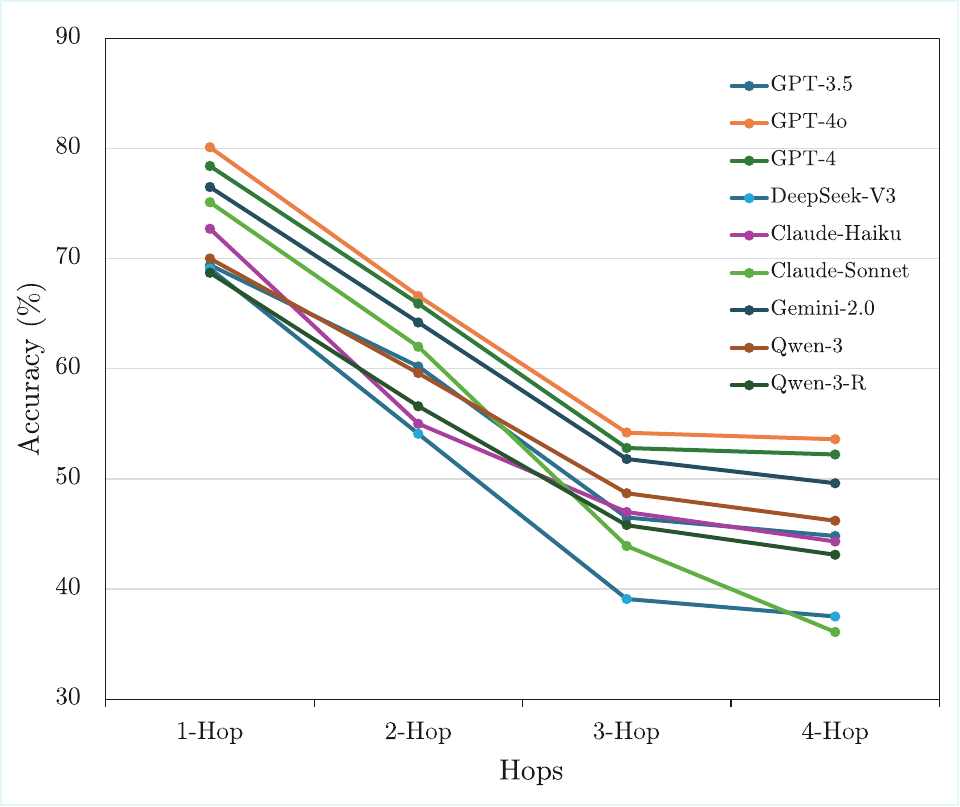}
    \caption{Average LLM accuracy across 1-hop to 4-hop tasks in all domains and question types.}
    \label{fig:bot_performance_hop}
\end{figure}

\textbf{Multi-hop questions amplify LLM challenges.} \methodname generates 1,000 multi-hop questions per domain and question type to assess complex reasoning. As depicted in Figure~\ref{fig:bot_performance_hop}, LLMs exhibit higher error rates on questions with more hops, with accuracy declining sharply from 1-hop to 2-hop questions and more gradually from 2-hop to 4-hop. This trend highlights the increased complexity of multi-hop reasoning, exacerbating factual inaccuracies, though the rate of accuracy decline slows with additional hops.

\begin{figure}[t]
    \centering
    \includegraphics[width=\linewidth]{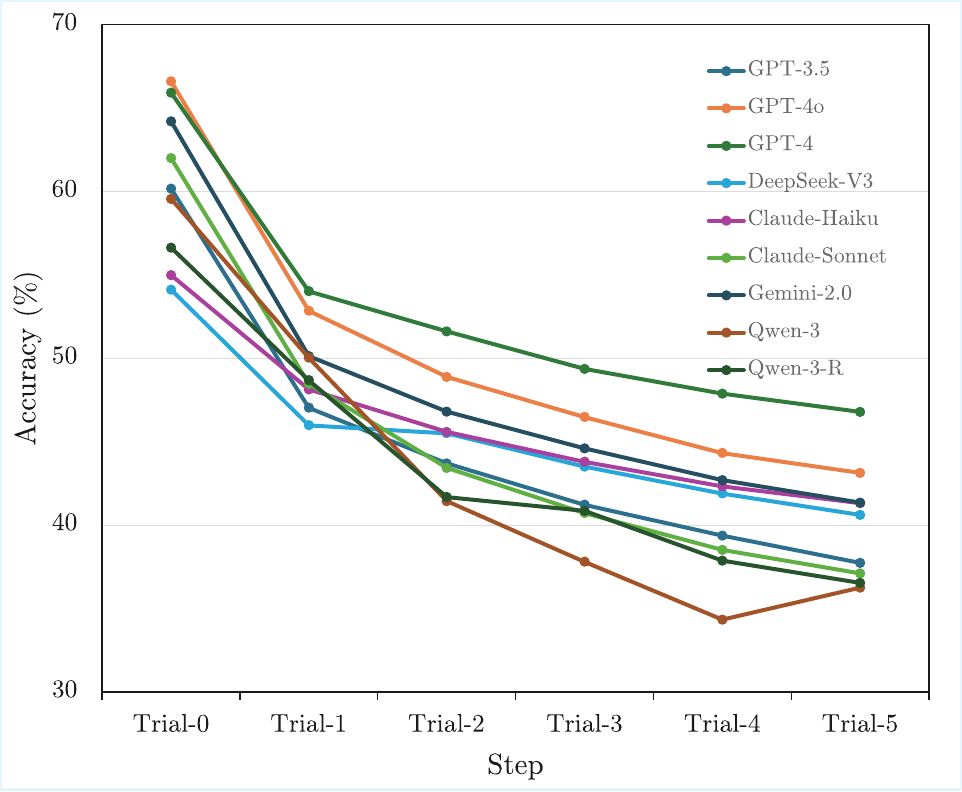}
    \caption{Average LLM accuracy trends across sequential trials for 2-hop questions.}
    \label{fig:llm-accuracy-trials-2hop}
\end{figure}

\begin{table}[t]
    \centering
    \caption{Coverage comparison between random questions and Trial 5 (cumulatively).}
    \resizebox{\linewidth}{!}{
    \begin{tabular}{l l c c}
        \toprule
        \textbf{Domain} & \textbf{Question Type} & \textbf{Trial 5 Coverage} & \textbf{Random Coverage} \\
        \midrule
        \multirow{3}{*}{Hum.} & Yes/No & 0.442 & 0.408 \\
                                    & MC & 0.480 & 0.366 \\
                                    & WH & 0.473 & 0.369 \\
        \midrule
        \multirow{3}{*}{Soc. Sci.} & Yes/No & 0.530 & 0.530 \\
                                        & MC & 0.523 & 0.551 \\
                                        & WH & 0.486 & 0.563 \\
        \midrule
        \multirow{3}{*}{STEM} & Yes/No & 0.445 & 0.333 \\
                              & MC & 0.442 & 0.250 \\
                              & WH & 0.437 & 0.280 \\
        \midrule
        \multicolumn{2}{c}{\textbf{Average}} & 0.473 & 0.406 \\
        \bottomrule
    \end{tabular}}
    \label{tab:coverage_comparison}
\end{table}
\begin{table}[th]
    \centering
    \caption{Median accuracy (absolute value) and median performance change (percentage vs. seed), aggregated across LLMs and question types, in different domains.}
    \resizebox{0.9\linewidth}{!}{
    \begin{tabular}{l c c c}
        \toprule
        \textbf{Trial} & \textbf{Hum. } & \textbf{Soc. Sci.} & \textbf{STEM} \\
        \midrule
        Seed & 0.712 (0.0\%) & 0.699 (-0.0\%) & 0.649 (-0.0\%) \\
        \midrule
        Trial 1 & 0.542 (-19.5\%) & 0.524 (-28.1\%) & 0.478 (-24.8\%) \\
        \midrule
        Trial 2 & 0.516 (-24.1\%) & 0.462 (-31.5\%) & 0.428 (-31.7\%) \\
        \midrule
        Trial 3 & 0.492 (-29.2\%) & 0.439 (-37.5\%) & 0.406 (-36.1\%) \\
        \midrule
        Trial 4 & 0.495 (-28.5\%) & 0.405 (-38.8\%) & 0.415 (-38.5\%) \\
        \midrule
        Trial 5 & 0.462 (-32.7\%) & 0.384 (-40.2\%) & 0.373 (-41.8\%) \\
        \bottomrule
    \end{tabular}}
    \label{tab:bot_performance_domains}
\end{table}

\subsection{Effectiveness of Iterative and Adaptive Question Generation}

This section assesses the efficacy of the iterative and adaptive question generation algorithm in enhancing exposure of factual error across Trials 0-5 in three question formats with fading color intensity, as shown in Figure~\ref{fig:heatmap_grouped}. Detailed results are provided in Appendix~\ref{sec:result_detail}.


\textbf{Targeted exposure of factual inaccuracies across domains.} \methodname’s iterative algorithm (Algorithm~\ref{alg:factprobe}) enhances the exposure of factual inaccuracies by generating targeted question sets that leverage LLM weaknesses identified from prior iterations, as demonstrated by consistent performance declines across the Humanity, Social Science, and STEM domains. As shown in Figure~\ref{fig:heatmap_grouped}, subsequent trials exhibit a significant reduction in LLM accuracy, driven by strategic triplet selection that probes deeper vulnerabilities. Table~\ref{tab:bot_performance_domains} details the median accuracies by Trial 5, with Social Science and STEM showing pronounced declines at 0.384 (-40.2\%) and 0.373 (-41.8\%), respectively, compared to Humanity’s milder drop at 0.462 (-32.7\%). These results suggest that LLMs struggle with the nuanced, complex questions in Social Science and the precision-demanding queries in STEM, while maintaining relatively stable performance in Humanity. This underscores HalluHunter’s ability to expose both factual and reasoning limitations across diverse domains through adaptive question generation.


\textbf{Robust error exposure in multi-hop scenarios.} The algorithm’s effectiveness extends to multi-hop questions requiring complex reasoning across multiple relations. For 2-hop questions, Figure~\ref{fig:llm-accuracy-trials-2hop} illustrates a consistent accuracy decline across trials, highlighting Algorithm~\ref{alg:factprobe}’s capability to probe deeper reasoning vulnerabilities. This underscores \methodname’s strength in revealing both factual and reasoning limitations in LLMs.

\textbf{Enhanced knowledge graph exploration.} To evaluate whether the iterative algorithm compromises exploration of the knowledge graph, we compare coverage between random question generation and Trial 5, both in total 5000 questions, as shown in Table~\ref{tab:coverage_comparison}. The results indicate that the algorithm either mostly maintains or increases coverage, ensuring that question generation remains diverse and covers a broad range of entities and relations in the knowledge graph.

\section{Related Work}
\label{sec-related}

The prevalent approach of factual evaluation involves reference-based techniques, which rely on benchmarks created through manual question design or test input labeling~\cite{Clark2019BoolQET, McCoy2019RightFT, Lin2021TruthfulQAMH, Talmor2019CommonsenseQAAQ, Laskar2023ASS}.
Despite their utility, these methods demand substantial human effort and are dependent on the development of comprehensive benchmarks.
\newcite{Lewis2021PAQ6M} is a pioneering work that automatically generates massive question-answer pairs to evaluate open-domain question-answering models. Recently, \newcite{Sun2023HeadtoTailHK} and \newcite{Omar2023ChatGPTVT} also adopted knowledge graphs to evaluate the factual correctness of LLMs. However, their scope of question types and topics, testbed models, and evaluation methods are limited, compared with \methodname.

Additionally, static benchmarks are prone to data contamination, posing challenges in accurately evaluating LLMs and efficiently identifying errors.
Recent advancements have seen the emergence of automatic test case generation, independent of manually pre-annotated labels~\cite{Chen2021TestingYQ,Liu2022QATestAU, Shen2022NaturalTG}.
However, these techniques still depend on existing benchmarks for question formulation, limiting the breadth of test case generation.

Although automated question generation from knowledge graphs (KGs) or LLMs reduces human annotation costs, the resulting questions are often generic and fail to target specific LLM weaknesses effectively. Recent approaches using LLMs to autonomously expose errors~\cite{chen2024llmQAiterate, cheng2024autodetect} rely heavily on LLMs for the evaluation pipeline, risking the propagation of inherent model biases and limiting test case diversity and grounding. In contrast, \methodname employs an iterative and adaptive algorithm that leverages a structured KG and embedding models to generate targeted, objective, and diverse test cases, ensuring effective uncovering of LLM weaknesses.

To sum up, our approach differs significantly in several key aspects:
(1) \textbf{Scope}: \methodname offers a more extensive scope, capable of generating three types of questions on any topic, as opposed to their method which is restricted to probing specific relationships, like ``place of birth'' or ``date of birth'', in a cloze-style format.
(2) \textbf{Testbed}: \methodname evaluates nine leading LLMs, while most of the previous works focused solely on the traditional models.
(3) \textbf{Testing Logic}: \methodname is an automated testing framework designed to dynamically generate a diverse array of test cases each time it is run, aiming to avoid data contamination.
(4) \textbf{Effective Generation}: \methodname’s algorithm uses performance-driven triplet selection to generate targeted, high-quality test cases, ensuring robust and effective uncovering of LLM weaknesses.

\section{Conclusion}

In this paper, we design and implement \methodname, an automated framework dedicated to uncovering factual errors in LLMs.
Distinct from previous approaches that depend on extensive human annotation or are prone to data contamination, \methodname leverages a structured KG to autonomously and iteratively generate a wide array of questions spanning various topics.
We conducted comprehensive evaluations using nine models.
Our empirical findings reveal that \methodname can successfully identify factual errors in LLMs.

\section*{Limitations}

This paper has two primary limitations that offer avenues for future research:
\begin{itemize}[leftmargin=*]
\item The effectiveness of \methodname is affected by the quality of the knowledge graph (\eg Wikidata), which is hard to guarantee. Please refer to our discussion in Appendix~\ref{sec-discuss}.
\item This paper does not provide any novel method to improve the factual correctness of LLMs. More effective and efficient methods are needed to further enhance the factual correctness of LLMs.
\end{itemize}

\bibliography{custom}

\clearpage

\appendix

\section{Threats to Validity} 
\label{sec-discuss}

The validity of this work may be subject to several potential threats.

The first concern is the reliance on NLP techniques employed by \methodname for error detection.
Given the inherent limitations of NLP methods, \methodname might generate false positives or overlook errors, resulting in false negatives.
This is particularly evident in scenarios where varying interpretations of correct responses to WH questions challenge accurate validation.
To mitigate this issue, we evaluated the efficacy of several prominent similarity methods, selecting the most effective one based on performance metrics. Additionally, we conducted human annotation to demonstrate that \methodname achieves high accuracy in error detection, as evidenced by the results.

The second threat is from the implementation of \methodname, which covers only one knowledge base, Wikidata.
Like any knowledge base, Wikidata is prone to factual inaccuracies or suffers from incomplete data, leading to sub-optimal question generation.
Additionally, it is vulnerable to issues like data leakage.
To address these two concerns, we adopt strategies respectively:
(1) \methodname is designed for flexibility, allowing easy substitution of Wikidata with alternative knowledge bases.
Incorporating multiple knowledge bases can enhance the robustness and quality of the generated questions.
(2) One advantage of Wikidata is the graph format for information storage, a method not extensively employed in training most LLMs despite its public availability.
Our primary contribution lies in the development of an automated testing framework.
This framework aims to minimize the human effort needed to identify factual inaccuracies within LLMs.
Essentially, \methodname flags potential errors, which are then subjected to further human analysis to assess their validity.

The third limitation of our study is the limited exploration of various LLMs during evaluation.
Our current analysis does not encompass a broad assessment of \methodname's performance across numerous systems.
To address this limitation, we focus on testing the most prevalent conversational LLMs and SOTA academic models developed by major corporations.
Future work, utilizing \methodname, could expand this scope to include additional commercial and research models, thereby enhancing the robustness of our findings.

\section{Illustration of The Retrieval Process of Fact Triplets.}

\begin{figure}[t]
    \centering
    \includegraphics[width=\linewidth]{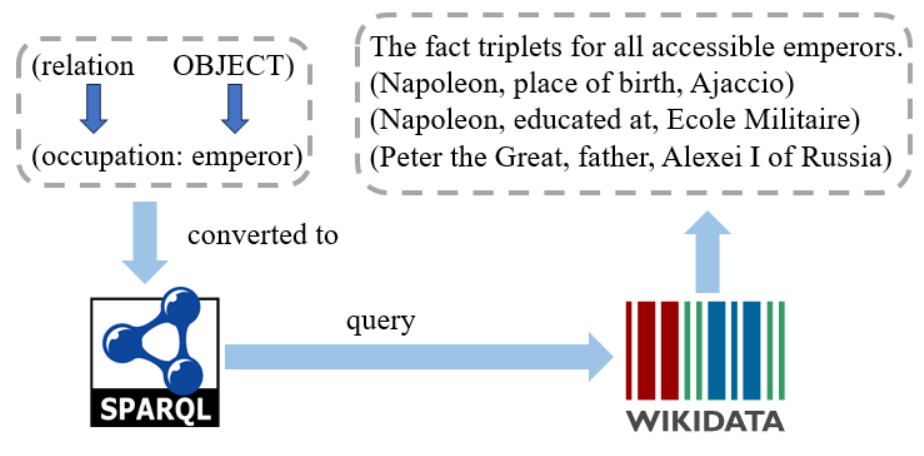}
    \caption{The retrieval process for fact triplets.}
    \label{fig:retrieval}
\end{figure}

A fact triplet is represented in the form of (SUBJECT, relation, OBJECT).
For instance, the triplet (USA, capital, Washington D.C.) denotes the fact that the capital of the USA is Washington D.C.
\methodname enables users to obtain fact triplets pertaining to specific topics.
As illustrated in Figure~\ref{fig:retrieval}, when a user expresses interest in the topic of emperors, \methodname proceeds to convert the ``{occupation: emperor}'' specification into a \texttt{SPARQL} query language\footnote{\url{https://www.wikidata.org/wiki/Wikidata:SPARQL_query_service}}, which is utilized for querying related triplets in Wikidata.
The resulting \texttt{SPARQL} query will retrieve all accessible fact triplets about emperors, including examples such as ``{Napoleon, place of birth, Ajaccio}'' and ``{Peter the Great, father, Alexei I of Russia}''. 

\section{Illustration of the Rule-Based Method for Question Generation}
\label{app:rule_based_explaination}

\begin{figure*}[t]
    \centering
    \includegraphics[width=\linewidth]{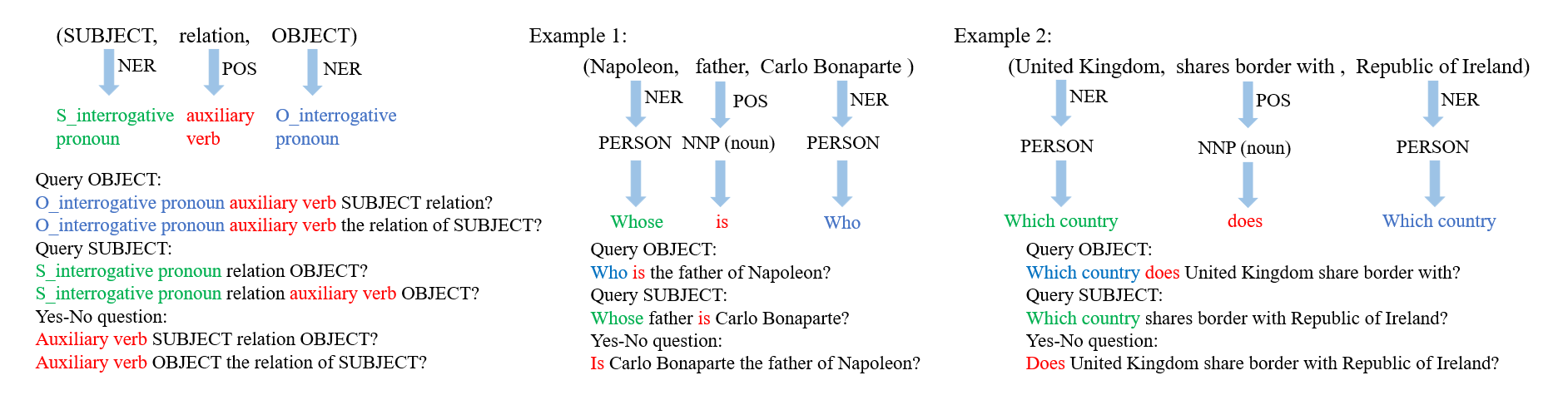}
    \caption{The proposed rule-based method for Question Generation.}
    \label{fig:rule-based}
\end{figure*}

\methodname utilizes a rule-based approach to generate questions from the constructed KG.
It can generate various types of questions, including different question types, \ie, Yes-No questions, MC questions and WH questions.
For each triplet in the constructed knowledge graph, \methodname converts it to the question form according to their POS and NER features.
Figure~\ref{fig:rule-based} shows some examples of \methodname's question generation method.

\section{Multi-Hop Question Construction}
\label{app:multi_hop_construction}

To address concerns about potential brittleness and ambiguity in multi-hop question construction, we provide a detailed specification of \methodname’s procedure, incorporating safeguards to prevent spurious chains and ensure clarity. The multi-hop generation is constrained to simple, two-hop chains of the form $(Entity_A, relation_X, Entity_B) \land (Entity_B, relation_Y, Entity_C)$, we turn them to $(Entity_A, \{relation_X, relation_Y\}, Entity_C)$. For example, to generate a question like "Where was Michelle Obama's spouse educated at?" (Table~\ref{tab:question_example}), \methodname retrieves the triplet (Michelle Obama, spouse, Barack Obama) and chains it with (Barack Obama, educated at, Harvard University). The intermediate entity ("spouse") is concatenated into a fluent subject ("Michelle Obama's spouse") using Part-of-Speech (PoS) analysis to determine the auxiliary verb ("was") and WH interrogative ("Where").

Generating WH questions for multi-hop scenarios poses a challenge, primarily concerning the assurance of answer uniqueness.
As an illustration, consider the triplet (Michelle Obama, {child, educated at}, Harvard Law School).
To assess the feasibility of employing this case for WH question generation, one \methodname must initially explore all of Michelle Obama's children.
In cases where Michelle Obama has multiple children, \methodname must then verify the educational background of each child.
Only when there exists a sole viable answer, the fact triplet becomes suitable for formulating the WH question.

\section{Costs of using \methodname}
The cost-efficiency of \methodname is a key advantage of its design. \methodname takes a few hours to generate tens of thousands of high-quality, multi-hop, multi-format questions. This automation eliminates the need for expensive crowd-sourced or expert annotations, which are commonly used in other benchmarks, and significantly reduces operational costs. Additionally, the framework avoids the reliance on LLM-driven question generation, which would incur additional API costs and potential biases. By iteratively re-mining the same KG to generate new, targeted questions that expose model weaknesses, \methodname maximizes the utility of the KG, further enhancing cost-efficiency. While the construction of a high-quality knowledge graph may require initial time and human effort, recent research has demonstrated the feasibility of automatic KG construction. For example, \cite{mo2025kggen} leverages LLMs to generate high-quality questions directly from automatically constructed KGs.

\section{Details of Selected Topics}

To ensure comprehensive evaluation, \methodname generate 1,000 questions per domain (Humanities, Social Sciences, STEM), with each domain comprising five topics (200 questions each). This distribution is maintained across trials for consistency. Table~\ref{tab:topic} lists the topics and example entities.

\begin{table}[t]
\centering
\caption{Selected topics for evaluation.}
\resizebox{1.0\linewidth}{!}{
\begin{tabular}{l l r}
\toprule
\bf Domain & \bf Topic  & \bf Example\\
\midrule
\multirow{3}{*}{Hum.}  &  Art &  Monna Lisa\\
&    History & World War II \\
& Philosophy & Plato \\
\hline
\multirow{3}{*}{Soc. Sci} & Psychology & Sigmund Freud \\
& Sociology &  Elitism \\
& Geography &  Earthquake  \\
\hline
\multirow{3}{*}{STEM} & Mathematics & Pythagorean theorem \\
& Physics & Planck constant\\
& Computer Science & Dijkstra's algorithm \\
\bottomrule
\end{tabular}}
\label{tab:topic}
\end{table}

\section{Preliminary Experiments}
\label{sec:prelim}
In this section, we conducted an initial experiment to validate our choice of employing a rule-based method for question generation as opposed to directly instructing ChatGPT(gpt-3.5-turbo-0613) to craft questions from fact triplets.
Additionally, we conducted experiments to assess the effectiveness of our grammar-checking and polishing modules.
We also meticulously investigated the comparison of five evaluation metrics.

\subsection{Prompts for ChatGPT}
To clarify, no prompt tuning was performed to optimize performance for ChatGPT experiments. Below, we provide three types of prompts for two question types (Yes-No and WH/MC) and two entity cases (asking about Entity A or B in relation (A, relation, B)) in MC and WH questions, presented in Table~\ref{prompts:yesno} and Table~\ref{prompts:Wh}.

\begin{table}[ht!]
\centering
\scriptsize
\caption{Yes/No Question template drafting.}
\begin{tabular}{p{0.45\textwidth}}
\toprule
Your are now given a triplet in format of <A, relation, B>, and your task is to create a yes/no question template for asking whether the relation holds between A and B. In crafting the template, you should replace entity A and entity B with <Subject> and <Object>, in order to build template for certain relation. \\\\

\#Example 1\\
Triplet: <Barack Obama, born in, Hawaii> \\
Question: Was <Subject> born in <Object>? \\\\

\#Example 2\\
Triplet: <Albert Einstein, profession, physicist> \\
Question: Was <Subject> a <Object>? 
\\\\

Now, create a question for the following triplet: \\
Triplet: \{SAMPLES\}\\
Question:  \\
\bottomrule
\end{tabular}
\label{prompts:yesno}

\end{table}

\begin{table}[ht!]
\centering
\scriptsize
\caption{Wh Question template drafting.}
\begin{tabular}{p{0.45\textwidth}}
\toprule
You are now given a triplet in format of <A, relation, B>, and your task is to create a wh- question template for asking Object B of the triplet. In crafting the template, you should replace entity A with <Subject> A, in order to build template for certain relation. 
 \\\\

\#Example 1\\
Triplet: <Barack Obama, born in, Hawaii> \\
Question: Where was <Subject> born? \\\\

\#Example 2\\
Triplet: <Albert Einstein, profession, physicist> \\
Question: What is the profession of <Subject>? 
\\\\

Now, create a question for the following triplet: \\
Triplet: \{SAMPLES\}\\
Question:  \\
\bottomrule
\end{tabular}
\label{prompts:Wh}

\end{table}

\subsection{Can ChatGPT outperform the proposed rule-based methods on question generation?}
\label{humanano}
Given the capabilities of ChatGPT, an alternative approach for generating questions from fact triplets involves instructing ChatGPT to generate the desired questions based on the extracted fact triplets.
To verify the viability of this approach, we prompted ChatGPT to generate 200 questions from fact triplets and compared the results with the rule-based method we proposed.
Subsequently, we enlisted the assistance of three annotators, each holding a Bachelor's degree or higher and proficient in English, to independently evaluate the quality of the generated questions.
Any discrepancies in their assessments were resolved through discussion.
The results indicate that while ChatGPT is capable of producing some high-quality questions from fact triplets, it may occasionally deviate from our instructions, introducing unreliability.
Among the 200 questions generated, the annotators found that 26 did not align with our expectations.
On the other hand, despite introducing some grammatical errors, the rule-based method produced questions where 98.5\% adhered to the intended semantic meaning.

\subsection{Are the modules of grammar-checking and rephrasing effective?}
In order to address potential grammatical errors introduced by rule-based approaches in question generation, we have implemented and compared two modules within our method.
The first approach incorporates the use of a grammar checker API to filter out questions exhibiting grammatical errors.
Following this filtering process, the remaining questions maintain a high level of quality.
However, this approach has a drawback as it tends to be overly sensitive, leading to the elimination of approximately 50\% of all generated questions, resulting in a notable false positive rate.
The second alternative entails instructing ChatGPT to paraphrase the generated questions, thereby rectifying grammatical errors and enhancing the natural sound of the questions.
Our hired annotators observed that by directly leveraging ChatGPT for paraphrasing, the question formats became more diverse, and all 48 questions initially containing grammar errors were successfully corrected.
The final results demonstrate that relying solely on the rewriting approach yields better overall performance.
Thus, \methodname adopts the rewriting powered by ChatGPT to obtain fluent test cases.

\subsection{Which similarity metric performs the best?}
Due to the diverse nature of the responses generated for WH questions, utilizing a straightforward exact match criterion is not sufficient for addressing these variations effectively. As a result, we compare five distinct evaluation methods described in Section~\ref{sec-error-indentify}. The objective is to identify the most effective method that yields satisfactory results. To conduct the evaluation, we randomly selected 500 questions along with their corresponding generated responses from LLMs and the ground-truth answers. Subsequently, the recruited three annotators are required to annotate whether the generated response matches the ground-truth answers. Finally, we obtained 238 cases that the responses are annotated as not aligned with the ground truth. Then, we use the annotated 500 data as a benchmark to evaluate the performance of the five matching methods. The results are shown in Table~\ref{tab:match_performance}, demonstrating that the sentence transformer method exhibits the most promising performance, with the highest F1 score of 87

It is important to note that this F1 score specifically applies to WH questions, which account for ~33\% of our benchmark (1,000 out of 3,000 questions per trial per domain). Other question types, such as Yes-No and Multiple-Choice, are evaluated deterministically via exact string matching, achieving 100\% evaluation accuracy. Thus, while WH questions inherently pose challenges due to answer variations (e.g., "Great Britain" vs. "United Kingdom" vs. "UK"), we consider the sentence transformer’s F1 score of 87\% reasonably high, given these complexities. In this context, we respectfully argue that our evaluation is both valuable and reliable.

Additionally, we emphasize that the sentence transformer method minimizes false negatives, which supports the reliability of our evaluation. As shown in Table~\ref{tab:match_performance}, the sentence transformer achieves relatively high precision—indicating few false positives—and exceptionally high recall, meaning nearly all true errors are captured. This high recall aligns with our primary focus on the reliability and quality of the identified set of factual inaccuracies. While there is an inherent trade-off between precision and recall, we prioritize recall to ensure that the identified error set is comprehensive and trustworthy for analyzing model weaknesses. Consequently, we consider a small number of missed errors acceptable, further supporting the use of sentence transformers as the matching metric for WH questions.

Thus, \methodname adopts the sentence transformer as the matching metric for WH questions.

\begin{table}[t]
\centering
\caption{Performance of different evaluation methods.}
\resizebox{0.49\textwidth}{!}{
\begin{tabular}{l r r r}
\toprule
\bf Evaluation Method & \bf Precision  & \bf Recall & \bf F1\\
\midrule
Levenshtein distance &  72.6 & 99.2 & 83.8\\
1-grams &   61.7 & 100 & 76.3\\
Word embedding & 72.8 & 91.2 & 81.0 \\
Sentence transformer & 78.2 & 97.9 & 87.0\\
ChatGPT & 100 & 65.5 & 79.2 \\
\bottomrule
\end{tabular}}
\label{tab:match_performance}
\end{table}

\section{Prompt Design for LLM Evaluation}
\label{sec:prompt_detail}

To ensure precise and consistent evaluation of Large Language Models (LLMs), \methodname employs carefully crafted prompts tailored to each question type, minimizing ambiguity and focusing on direct responses to assess factual accuracy. The specific prompts used are as follows:

\begin{itemize}[leftmargin=*]
    \item \textbf{Yes-No Questions}: ``The following question's topic is about \texttt{TOPIC}. Only answer `Yes' or `No', and do not provide an explanation.''
    \item \textbf{Multiple-Choice Questions}: ``The following question's topic is about \texttt{TOPIC}. Choose the only correct option from (`A', `B', `C', or `D') and do not provide an explanation.''
    \item \textbf{WH Questions}: ``The following question's topic is about \texttt{TOPIC}. Provide the answer in `phrase' or `word' format only. Do not provide an explanation or use a sentence.''
\end{itemize}

It is important to note that this framework does not allow the model to abstain or refrain from providing an answer. Abstention benchmarks like SimpleQA \cite{wei2024SimpleQa} are designed to evaluate whether models know when they don’t know, measuring calibration and epistemic uncertainty. In contrast, \methodname focuses on whether models correctly recall and reason over facts, measuring parametric knowledge coverage and effectively exposing factual inaccuracies in different models. Both dimensions are crucial for building trustworthy LLMs. However, while abstention benchmarks prioritize precision by focusing on reducing false positives, our study prioritizes recall by targeting the identification of factual inaccuracies.

From this perspective, our study regards abstention as incorrect, as the goal is to rigorously test the model’s factual correctness. Nevertheless, we emphasize that in real downstream applications, abstention is often preferable to unsupported guessing when reliability and safety are the primary concern.


\section{Reproducibility Details}
\label{sec:reproducibility}

To enhance reproducibility, we clarify key configurations and hyperparameters for \methodname. The knowledge graph is constructed from Wikidata, with each domain containing over 500,000 triplets and more than 10,000 distinct entities. Random selection experiments use a fixed seed for deterministic outcomes. The iterative algorithm (Algorithm~\ref{alg:factprobe}) employs the following hyperparameters: an explore constant $e = 0.2$ balances exploitation of low-accuracy relations ($R_{low}$) with exploration of new relations; a low-accuracy constant $a = 0.4$ targets the 40th percentile of relation accuracies; and a top-$k$ parameter $k = 10$ limits the QuatE embedding model (PyKEEN) to the 10 most similar entities for distractor generation. For answer assessment, the sentence transformer similarity threshold is set to 0.6, with performance detailed in Table~\ref{tab:match_performance}. Experiments were conducted on an A100 GPU (40GB VRAM), with API calls to nine LLMs (e.g., GPT-4o, Claude-3.5-Haiku) costing approximately \$400 USD.

\section{Comprehensive Evaluation Results}
\label{sec:result_detail}

This section details \methodname’s evaluation of LLM factual accuracy across single-hop, 2-hop, and multi-hop questions, highlighting the iterative algorithm’s effectiveness in exposing inaccuracies.

\subsection{Multi-Hop Question Performance}
\label{sec:multi_hop_results}

Table~\ref{tab:bot-performance-multi-detail} shows LLM accuracy on multi-hop questions (1-4 hops) across Humanities, Social Sciences, and STEM for Yes-No and Multiple-Choice questions. Each trial includes 1,000 unique questions per domain and question type. Accuracy decreases with increasing hops, with Multiple-Choice questions posing greater challenges, especially in STEM.

\subsection{Single-Hop Question Performance}
\label{sec:single_hop_results}

Figure~\ref{fig:heatmap_grouped}  employs three distinct color gradients to differentiate between question types: orange represents Yes/No questions, blue signifies Multiple-Choice (MC) questions, and green corresponds to WH questions. Within each gradient, the intensity of the color reflects the models' performance across sequential trials, with darker shades indicating higher accuracy and better performance, while lighter shades, closer to white, denote lower accuracy. This gradation effectively highlights the impact of our iterative algorithm in exposing factual inaccuracies.

Table~\ref{tab:llm-accuracy-trials} presents LLM accuracy on single-hop questions across three domains and question types (Yes-No, Multiple-Choice, WH). Each trial includes 1,000 unique questions. Algorithm~\ref{alg:factprobe} reduces accuracy in later trials (e.g., GPT-4o’s accuracy drops from 84.4\% to 65.8\% for Yes-No and 82.9\% to 54.1\% for Multiple-Choice), with WH questions showing the lowest average accuracy (37.4\%). 

\subsection{2-Hop Question Performance}
\label{sec:2hop_results}

Table~\ref{tab:llm-accuracy-trials-2hop} reports LLM accuracy on 2-hop questions across domains and question types. The iterative algorithm significantly lowers accuracy in subsequent trials, highlighting LLMs’ challenges with sequential reasoning.

\subsection{Validation of Factual Errors}
\label{humanano2}

To investigate whether \methodname reveals novel weaknesses in LLMs beyond reformulating known deficiencies, we conducted a comparative analysis of its performance against established benchmarks. Figure~\ref{fig:bot_performance_hop} illustrates LLM performance on single-hop questions generated through random triplet selection, representing the baseline (Trial 0) without \methodname’s iterative probing mechanism. These results align closely with the baseline accuracies reported in Table~\ref{tab:llm-accuracy-trials}, where models such as GPT-4o, GPT-4, and Gemini-2.0 achieve accuracies ranging from 65\% to 70\% across domains.

However, the application of \methodname’s iterative and adaptive question generation algorithm, as detailed in Algorithm~\ref{alg:factprobe}, significantly amplifies the exposure of LLM vulnerabilities. As shown in Tables~\ref{tab:llm-accuracy-trials} and~\ref{tab:llm-accuracy-trials-2hop}, accuracies across Trials 1–5 decline markedly, with most LLMs dropping to approximately 50\%, 40\%, and 10\% for Yes-No, Multiple-Choice (MC), and WH questions, respectively. This pronounced reduction, which intensifies with successive trials, demonstrates \methodname’s efficacy in systematically revealing error-prone areas that challenge LLMs’ factual recall and reasoning capabilities.

In contrast, benchmarks such as Head-to-Tail~\cite{Sun2023HeadtoTailHK} report GPT-4o achieving approximately 40\% accuracy on WH questions in open-domain settings. \methodname’s iterative probing further exacerbates this decline, projecting accuracies as low as 10\% by Trial 5 across all domains, as evidenced in Table~\ref{tab:llm-accuracy-trials}. This substantial degradation highlights \methodname’s unique ability to uncover novel failure modes that conventional benchmarks may overlook, thereby providing deeper insights into LLMs’ factual limitations and enhancing the evaluation of their robustness.

\subsection{Error Pattern Analysis}
\begin{table}[h]
\centering
\caption{Error Patterns of different models in specific relations}
\resizebox{0.49\textwidth}{!}{
\begin{tabular}{l l l r}
\toprule
\bf Models & \bf Domains  & \bf Relations  & \bf Accuracy \\
\midrule
\multirow{2}{*}{GPT-4o} & \multirow{2}{*}{Physics} & Binding energy & 0.258 (74/287)\\
& & Mass excess & 0.237 (69/291) \\
\midrule
\multirow{2}{*}{Claude Haiku} & Mathematics  & Prime factor & 0.313 (260/831)\\
& Biology & Chromosome & 0.237 (69/291) \\
\midrule
\multirow{2}{*}{Gemini} & Physics & Mass excess & 0.241 (67/278) \\
 & Computer Science & Operating system & 0.308 (104/367) \\
\midrule
DeepSeek & Economics & Central bank/issuer & 0.253 (39/154) \\
\bottomrule
\end{tabular}}
\label{tab:error_pattern}
\end{table}

We have conducted a detailed examination of the complete six-trial dataset, comprising 6,000 questions. This analysis reveals several key observations regarding topic-level trends and model-specific blind spots. For instance, GPT-4o, the overall strongest model in our evaluation, exhibits clear weaknesses in nuclear and particle physics concepts. Relations such as “binding energy” and “mass excess” yield accuracies of only 0.258 and 0.237, respectively. In contrast, GPT-4o performs near ceiling on biomedical associations, achieving an accuracy of 0.778 across questions related to “genetic association” These results highlight GPT-4o’s blind spots in physics-related domains despite its exceptional performance in biology, suggesting that its training data or reasoning capabilities may be biased toward biomedical topics.

Similarly, Claude-3.5-Haiku demonstrates a reproducible weakness in elementary number-theory reasoning. For questions related to “prime factor,” Claude-3.5-Haiku achieves an accuracy of only 0.313 across 831 questions, whereas other models such as Gemini-2.0 and GPT-4o consistently perform 0.60 on the exact same domain questions. This discrepancy underscores a specific limitation of the Claude-3.5-Haiku in mathematical reasoning, which our adaptive rounds systematically surface. These cases are similar in Cytology domains. These findings not only validate the effectiveness of our framework in exposing model-specific weaknesses but also provide actionable insights into the areas where LLMs require further improvement. Future researchers can use our framework to improve their models based on the performance analysis.

\subsection{Coverage Analysis}
Our algorithm~\ref{alg:factprobe} strengthens exploitation through targeted probing of model weaknesses, without sacrificing coverage compared to random sampling. Specifically, exploitation focuses only on queries where the model fails. 
Furthermore, the model does not get stuck in specific regions of the KG. We have calculated the coverage metric across multiple trials, which shows a steady increase from the initial seed trial to the final trial, instead of showing a sign of stuck. Table~\ref{tab:llm-coveragescore-trials} reports the coverage scores of our algorithm for each trial, domain, and LLM. A higher coverage score indicates that the algorithm explores and covers more of the knowledge graph effectively.

\subsection{Exploration-Exploitation Tradeoff Analysis}
The adaptive algorithm in HalluHunter balances exploitation of identified weaknesses against exploration of new areas of the knowledge graph. As shown in Algorithm \ref{alg:factprobe}, this tradeoff is controlled by two hyperparameters: the exploration constant $e$, which determines how often the algorithm prioritizes low-accuracy relations, and the low-accuracy threshold $a$, which defines what counts as a weak relation. 

To better characterize the exploration-exploitation (EE) dynamics, we conduct a sensitivity analysis by varying one hyperparameter at a time. Specifically, we vary $a \in \{0.3, 0.4, 0.5\}$ with $e=0.2$ fixed, and vary $e \in \{0.1, 0.2, 0.3\}$ with $a=0.4$ fixed. For each setting, we report the average accuracy and average coverage across question types and domains.

\paragraph{Sensitivity to the low-accuracy threshold $a$.}
Table~\ref{tab:ee_a} shows that a lower threshold ($a=0.3$) induces more aggressive exploitation of weak relations. This yields slightly lower final accuracy, but also noticeably lower coverage, suggesting that the algorithm focuses too narrowly on a smaller subset of relations. In contrast, a larger threshold ($a=0.5$) relaxes exploitation, resulting smaller initial accuracy drops and stabilizing at higher values, with a comparable coverage to $a=0.4$. This highlights the tradeoff that stricter constants enhance short-term error exposure but risk coverage stagnation, while looser ones promote broader exploration at the cost of slower vulnerability detection.

\begin{table*}[t]
\centering
\small
\caption{Sensitivity analysis for the low-accuracy threshold $a$ with $e=0.2$ fixed.}
\label{tab:ee_a}
\begin{tabular}{lcccccc}
\toprule
Trial & \multicolumn{2}{c}{$a=0.3$} & \multicolumn{2}{c}{$a=0.4$} & \multicolumn{2}{c}{$a=0.5$} \\
\cmidrule(lr){2-3} \cmidrule(lr){4-5} \cmidrule(lr){6-7}
& Accuracy & Coverage & Accuracy & Coverage & Accuracy & Coverage \\
\midrule
Seed    & 0.605 & 0.314 & 0.605 & 0.314 & 0.605 & 0.314 \\
1       & 0.460 & 0.307 & 0.481 & 0.326 & 0.522 & 0.346 \\
2       & 0.443 & 0.370 & 0.444 & 0.391 & 0.507 & 0.386 \\
3       & 0.420 & 0.397 & 0.446 & 0.425 & 0.495 & 0.409 \\
4       & 0.404 & 0.411 & 0.424 & 0.449 & 0.489 & 0.448 \\
5       & 0.371 & 0.417 & 0.373 & 0.471 & 0.450 & 0.468 \\
\bottomrule
\end{tabular}
\end{table*}

\paragraph{Sensitivity to the exploration constant $e$.}
Table~\ref{tab:ee_e} shows that $e=0.2$ produces the steepest and most consistent decline in accuracy, indicating the strongest exposure of factual weaknesses. By comparison, $e=0.1$ and $e=0.3$ lead to more modest drops. Although the final coverage of $e=0.2$ is similar to that of $e=0.3$, the setting $e=0.1$ exhibits less stable behavior across trials. 

\begin{table*}[t]
\centering
\small
\caption{Sensitivity analysis for the exploration constant $e$ with $a=0.4$ fixed.}
\label{tab:ee_e}
\begin{tabular}{lcccccc}
\toprule
Trial & \multicolumn{2}{c}{$e=0.1$} & \multicolumn{2}{c}{$e=0.2$} & \multicolumn{2}{c}{$e=0.3$} \\
\cmidrule(lr){2-3} \cmidrule(lr){4-5} \cmidrule(lr){6-7}
& Accuracy & Coverage & Accuracy & Coverage & Accuracy & Coverage \\
\midrule
Seed    & 0.605 & 0.314 & 0.605 & 0.314 & 0.605 & 0.314 \\
1       & 0.450 & 0.348 & 0.481 & 0.326 & 0.498 & 0.315 \\
2       & 0.436 & 0.380 & 0.444 & 0.391 & 0.468 & 0.378 \\
3       & 0.462 & 0.408 & 0.446 & 0.425 & 0.435 & 0.413 \\
4       & 0.455 & 0.452 & 0.424 & 0.449 & 0.429 & 0.435 \\
5       & 0.430 & 0.460 & 0.373 & 0.471 & 0.412 & 0.472 \\
\bottomrule
\end{tabular}
\end{table*}

\paragraph{How many evaluation points are needed?}
At the scale used in our experiments, each domain-specific KG subset contains roughly 500k-600k triplets and 10k-12k entities, and each iteration generates 1,000 new questions per question type and domain. The results in Tables~\ref{tab:ee_a} and \ref{tab:ee_e} show that the first two iterations quickly exploit initial weaknesses, producing sharp drops in accuracy but only modest coverage. Stopping too early therefore risks missing substantial vulnerabilities that emerge only after broader exploration.

In practice, we recommend a dynamic stopping criterion that they continue iterations until marginal growth in coverage diminishes or accuracy stabilizes per trial. This rule ensures efficient scaling, balances the EE tradeoff, and prevents unnecessary over-sampling.

\section{Statistical Analysis of Iterative Algorithm Effectiveness}
\label{sec:stat_analysis}

To address concerns regarding the statistical rigor of \methodname’s evaluation and the need for a more measured narrative on its impact, we conducted additional statistical analysis to quantify the effectiveness of the iterative algorithm (Algorithm~\ref{alg:factprobe}). Specifically, we performed linear regression on the average accuracy of the nine evaluated LLMs (GPT-3.5-Turbo, GPT-4-Turbo, GPT-4o, DeepSeek-V3, Claude-Sonnet-4, Claude-3.5-Haiku, Gemini-2.0, Qwen-3, Qwen-3-Reasoning) across six iterations (Trials 0–5, with Trial 0 as the initial random question generation). The analysis was separated for single-hop and multi-hop questions to assess the algorithm’s ability to progressively uncover LLM vulnerabilities.

For single-hop questions, linear regression on the average accuracy across iterations yielded a p-value of 0.031, indicating a statistically significant negative trend (declining accuracy as iterations target model weaknesses). For multi-hop questions, the p-value was 0.01, further confirming the algorithm’s effectiveness in exposing factual inaccuracies through iterative probing. These results, detailed in Sections~\ref{sec:single_hop_results} and~\ref{sec:multi_hop_results}, are supported by corresponding graphs in the revised paper, enhancing analytical transparency. Claims about \methodname’s impact are now confined to its demonstrated ability to iteratively uncover LLM vulnerabilities through automated question-answering probing, avoiding overgeneralization.

\begin{table*}[t]
  \centering
  \caption{Factual accuracy of different LLMs on
multi-hop questions.}
  \resizebox{0.6\textwidth}{!}{
  \begin{tabular}{ll ll cccc}
    \toprule
    \multirow{2}{*}{LLM} & \multirow{2}{*}{Domain} & \multirow{2}{*}{Question Type} & \multicolumn{4}{c}{Hop} \\
    \cmidrule(lr){4-7}
    & & & 1-hop & 2-hop & 3-hop & 4-hop \\
    \midrule
    \multirow{6}{*}{GPT-3.5} 
    & \multirow{2}{*}{Hum.} & Yes/No & 72.2 & 68.7 & 54.9 & 51.4 \\
    & & MC & 68.6 & 64.2 & 46.0 & 42.4 \\
    & \multirow{2}{*}{Soc. Sci.} & Yes/No & 74.1 & 66.2 & 53.9 & 50.0 \\
    & & MC & 67.0 & 50.3 & 39.2 & 43.8 \\
    & \multirow{2}{*}{STEM} & Yes/No & 73.9 & 66.4 & 53.8 & 51.4 \\
    & & MC & 60.5 & 45.2 & 31.3 & 29.5 \\
    \midrule
    \multirow{6}{*}{GPT-4o} 
    & \multirow{2}{*}{Hum.} & Yes/No & 84.4 & 74.4 & 62.9 & 60.6 \\
    & & MC & 82.9 & 74.8 & 56.7 & 50.6 \\
    & \multirow{2}{*}{Soc. Sci.} & Yes/No & 82.1 & 68.8 & 59.7 & 64.9 \\
    & & MC & 79.2 & 61.5 & 54.1 & 55.4 \\
    & \multirow{2}{*}{STEM} & Yes/No & 79.4 & 70.5 & 57.0 & 56.4 \\
    & & MC & 72.6 & 49.6 & 34.6 & 33.4 \\
    \midrule
    \multirow{6}{*}{GPT-4} 
    & \multirow{2}{*}{Hum.} & Yes/No & 81.0 & 73.4 & 61.2 & 60.9 \\
    & & MC & 79.2 & 72.4 & 55.0 & 47.7 \\
    & \multirow{2}{*}{Soc. Sci.} & Yes/No & 80.7 & 68.3 & 58.7 & 65.4 \\
    & & MC & 77.6 & 62.0 & 53.0 & 53.8 \\
    & \multirow{2}{*}{STEM} & Yes/No & 81.0 & 69.2 & 56.5 & 53.9 \\
    & & MC & 70.9 & 50.2 & 32.3 & 31.3 \\
    \midrule
    \multirow{6}{*}{DeepSeek-V3} 
    & \multirow{2}{*}{Hum.} & Yes/No & 76.2 & 67.7 & 56.0 & 52.1 \\
    & & MC & 65.0 & 56.3 & 31.5 & 28.0 \\
    & \multirow{2}{*}{Soc. Sci.} & Yes/No & 75.6 & 66.2 & 55.6 & 51.0 \\
    & & MC & 65.0 & 45.2 & 24.5 & 26.5 \\
    & \multirow{2}{*}{STEM} & Yes/No & 77.2 & 62.4 & 53.0 & 51.5 \\
    & & MC & 55.5 & 26.9 & 14.1 & 16.0 \\
    \midrule
    \multirow{6}{*}{Claude-3.5-Haiku} 
    & \multirow{2}{*}{Hum.} & Yes/No & 76.9 & 70.5 & 60.4 & 55.6 \\
    & & MC & 71.2 & 55.7 & 39.7 & 33.6 \\
    & \multirow{2}{*}{Soc. Sci.} & Yes/No & 78.9 & 55.4 & 60.0 & 60.0 \\
    & & MC & 69.9 & 43.8 & 36.0 & 34.9 \\
    & \multirow{2}{*}{STEM} & Yes/No & 76.5 & 68.1 & 56.8 & 54.9 \\
    & & MC & 62.9 & 36.4 & 29.3 & 27.0 \\
    \midrule
    \multirow{6}{*}{Claude-Sonnet-4.0} 
    & \multirow{2}{*}{Hum.} & Yes/No & 78.1 & 70.2 & 60.1 & 52.7 \\
    & & MC & 75.3 & 68.5 & 33.4 & 19.3 \\
    & \multirow{2}{*}{Soc. Sci.} & Yes/No & 79.0 & 70.2 & 59.9 & 55.9 \\
    & & MC & 74.3 & 56.8 & 31.4 & 22.6 \\
    & \multirow{2}{*}{STEM} & Yes/No & 78.9 & 66.1 & 54.8 & 51.6 \\
    & & MC & 65.2 & 40.2 & 23.5 & 14.2 \\
    \midrule
    \multirow{6}{*}{Gemini-2.0} 
    & \multirow{2}{*}{Hum.} & Yes/No & 81.3 & 73.1 & 62.0 & 57.8 \\
    & & MC & 77.7 & 70.1 & 52.1 & 45.9 \\
    & \multirow{2}{*}{Soc. Sci.} & Yes/No & 78.9 & 67.6 & 57.4 & 56.2 \\
    & & MC & 75.6 & 59.4 & 49.1 & 51.0 \\
    & \multirow{2}{*}{STEM} & Yes/No & 76.0 & 68.5 & 55.1 & 54.4 \\
    & & MC & 69.6 & 46.5 & 35.2 & 32.1 \\
    \midrule
    \multirow{6}{*}{Qwen-3} 
    & \multirow{2}{*}{Hum.} & Yes/No & 74.4 & 69.3 & 59.0 & 55.6 \\
    & & MC & 67.4 & 63.4 & 45.1 & 42.0 \\
    & \multirow{2}{*}{Soc. Sci.} & Yes/No & 72.9 & 65.5 & 59.8 & 56.2 \\
    & & MC & 66.0 & 51.6 & 43.4 & 43.1 \\
    & \multirow{2}{*}{STEM} & Yes/No & 74.5 & 65.0 & 53.9 & 52.1 \\
    & & MC & 64.9 & 42.5 & 31.0 & 28.4 \\
    \midrule
    \multirow{6}{*}{Qwen-3-Reasoning} 
    & \multirow{2}{*}{Hum.} & Yes/No & 72.8 & 67.6 & 56.1 & 54.1 \\
    & & MC & 63.9 & 59.9 & 41.0 & 34.5 \\
    & \multirow{2}{*}{Soc. Sci.} & Yes/No & 75.1 & 65.8 & 57.6 & 52.0 \\
    & & MC & 62.6 & 44.0 & 37.3 & 38.3 \\
    & \multirow{2}{*}{STEM} & Yes/No & 76.6 & 61.2 & 52.9 & 51.6 \\
    & & MC & 61.2 & 41.3 & 30.1 & 28.3 \\
    \bottomrule
  \end{tabular}}
  \label{tab:bot-performance-multi-detail}
\end{table*}

\begin{table*}[h]
  \centering
  \caption{Factual accuracy of LLMs across multiple trials on single-hop questions.}
  \label{tab:llm-accuracy-trials}
  \resizebox{0.8\textwidth}{!}{
  \begin{tabular}{ll ccc ccc ccc}
    \toprule
    \multirow{2}{*}{LLM} & \multirow{2}{*}{Trial} & \multicolumn{3}{c}{Hum.} & \multicolumn{3}{c}{Soc. Sci.} & \multicolumn{3}{c}{STEM} \\
    \cmidrule(lr){3-5} \cmidrule(lr){6-8} \cmidrule(lr){9-11}
    & & Yes/No & MC & WH & Yes/No & MC & WH & Yes/No & MC & WH \\
    \midrule
    \multirow{6}{*}{GPT-3.5} 
    & Trial 0 & 72.2 & 68.6 & 45.3 & 74.1 & 67.0 & 44.3 & 73.9 & 60.5 & 35.3 \\
    & Trial 1 & 61.8 & 52.6 & 18.5 & 58.1 & 50.4 & 23.4 & 62.1 & 48.1 & 18.4 \\
    & Trial 2 & 57.5 & 49.6 & 12.5 & 52.8 & 44.2 & 21.1 & 55.1 & 44.4 & 15.8 \\
    & Trial 3 & 56.2 & 46.9 & 11.4 & 53.0 & 42.4 & 15.4 & 57.2 & 44.6 & 15.5 \\
    & Trial 4 & 55.3 & 45.5 & 9.8 & 51.9 & 39.9 & 14.4 & 55.4 & 42.4 & 11.1 \\
    & Trial 5 & 53.6 & 46.2 & 7.5 & 48.1 & 38.2 & 12.1 & 53.2 & 37.3 & 11.1 \\
    \midrule
    \multirow{6}{*}{GPT-4o} 
    & Trial 0 & 84.4 & 82.9 & 50.8 & 82.1 & 79.2 & 51.4 & 79.4 & 72.6 & 38.4 \\
    & Trial 1 & 71.0 & 68.3 & 20.4 & 74.5 & 65.2 & 21.4 & 71.0 & 61.3 & 21.1 \\
    & Trial 2 & 66.2 & 58.9 & 15.0 & 68.1 & 54.1 & 16.5 & 65.7 & 57.4 & 15.2 \\
    & Trial 3 & 67.3 & 56.4 & 12.0 & 67.4 & 54.7 & 14.9 & 65.7 & 54.9 & 15.1 \\
    & Trial 4 & 65.7 & 51.7 & 8.8 & 63.2 & 50.8 & 11.4 & 61.0 & 52.0 & 10.8 \\
    & Trial 5 & 65.8 & 54.1 & 9.1 & 63.6 & 51.5 & 10.6 & 61.2 & 50.5 & 10.3 \\
    \midrule
    \multirow{6}{*}{GPT-4} 
    & Trial 0 & 81.0 & 79.2 & 49.1 & 80.7 & 77.6 & 48.4 & 81.0 & 70.9 & 39.3 \\
    & Trial 1 & 69.0 & 67.1 & 25.4 & 63.4 & 57.2 & 24.2 & 61.9 & 54.1 & 18.4 \\
    & Trial 2 & 59.4 & 61.6 & 18.4 & 65.1 & 56.5 & 21.8 & 60.8 & 50.1 & 16.8 \\
    & Trial 3 & 61.3 & 61.1 & 16.0 & 59.2 & 49.6 & 20.8 & 61.9 & 48.2 & 14.6 \\
    & Trial 4 & 62.3 & 60.8 & 14.3 & 60.2 & 53.1 & 15.2 & 57.4 & 46.2 & 13.0 \\
    & Trial 5 & 61.0 & 57.5 & 14.4 & 57.4 & 49.1 & 14.5 & 55.0 & 45.8 & 9.8 \\
    \midrule
    \multirow{6}{*}{DeepSeek-V3} 
    & Trial 0 & 76.2 & 65.0 & 51.7 & 75.6 & 65.0 & 44.9 & 77.2 & 55.5 & 39.0 \\
    & Trial 1 & 63.1 & 52.3 & 23.6 & 61.8 & 43.4 & 20.1 & 59.7 & 37.5 & 20.0 \\
    & Trial 2 & 57.8 & 49.9 & 16.7 & 56.6 & 43.8 & 18.8 & 58.8 & 32.1 & 14.2 \\
    & Trial 3 & 52.8 & 49.2 & 11.3 & 58.3 & 41.5 & 17.1 & 55.0 & 28.4 & 13.9 \\
    & Trial 4 & 55.6 & 45.9 & 11.5 & 50.5 & 38.0 & 14.4 & 50.8 & 30.0 & 12.2 \\
    & Trial 5 & 52.3 & 43.2 & 10.1 & 51.6 & 37.9 & 10.9 & 51.1 & 29.6 & 10.8 \\
    \midrule
    \multirow{6}{*}{Claude-3.5-Haiku} 
    & Trial 0 & 76.9 & 71.2 & 45.5 & 78.9 & 69.9 & 45.4 & 76.5 & 62.9 & 34.1 \\
    & Trial 1 & 66.4 & 54.2 & 17.9 & 65.2 & 47.0 & 22.8 & 60.7 & 41.3 & 18.7 \\
    & Trial 2 & 62.1 & 49.1 & 15.3 & 60.6 & 42.7 & 18.6 & 57.4 & 40.0 & 15.9 \\
    & Trial 3 & 61.1 & 48.1 & 13.7 & 60.1 & 40.5 & 19.0 & 58.3 & 38.1 & 15.3 \\
    & Trial 4 & 57.4 & 45.4 & 11.8 & 59.4 & 37.0 & 16.1 & 54.4 & 34.7 & 13.0 \\
    & Trial 5 & 57.6 & 44.9 & 11.5 & 57.8 & 38.2 & 16.3 & 56.1 & 36.4 & 12.3 \\
    \midrule
    \multirow{6}{*}{Claude-Sonnet-4.0} 
    & Trial 0 & 78.1 & 75.3 & 52.8 & 79.0 & 74.3 & 50.2 & 78.9 & 65.2 & 44.3 \\
    & Trial 1 & 64.8 & 66.4 & 23.1 & 58.3 & 59.3 & 25.8 & 56.7 & 45.6 & 21.6 \\
    & Trial 2 & 63.0 & 62.8 & 16.2 & 52.7 & 56.7 & 21.6 & 56.6 & 34.4 & 15.9 \\
    & Trial 3 & 60.7 & 59.5 & 14.6 & 49.8 & 55.4 & 21.6 & 52.5 & 30.8 & 16.7 \\
    & Trial 4 & 61.0 & 58.6 & 14.0 & 48.6 & 47.6 & 17.4 & 54.4 & 26.0 & 13.5 \\
    & Trial 5 & 55.0 & 57.1 & 10.4 & 44.9 & 42.5 & 16.0 & 53.1 & 24.7 & 13.4 \\
    \midrule
    \multirow{6}{*}{Gemini-2.0} 
    & Trial 0 & 81.3 & 77.7 & 48.8 & 78.9 & 75.6 & 50.3 & 76.0 & 69.6 & 37.7 \\
    & Trial 1 & 65.7 & 63.4 & 21.2 & 62.8 & 64.4 & 24.3 & 61.3 & 52.0 & 20.1 \\
    & Trial 2 & 63.5 & 61.1 & 15.3 & 63.4 & 62.1 & 23.3 & 59.9 & 47.7 & 16.1 \\
    & Trial 3 & 62.7 & 60.5 & 12.0 & 62.4 & 57.8 & 22.2 & 59.0 & 43.5 & 14.7 \\
    & Trial 4 & 60.9 & 60.2 & 12.4 & 60.5 & 58.2 & 20.3 & 55.6 & 42.6 & 15.9 \\
    & Trial 5 & 62.4 & 55.2 & 11.6 & 59.7 & 50.0 & 23.7 & 56.5 & 41.6 & 11.3 \\
    \midrule
    \multirow{6}{*}{Qwen-3} 
    & Trial 0 & 74.4 & 67.4 & 42.0 & 72.9 & 66.0 & 41.2 & 74.5 & 64.9 & 35.7 \\
    & Trial 1 & 66.8 & 52.3 & 18.9 & 62.3 & 52.4 & 18.1 & 63.2 & 47.8 & 16.2 \\
    & Trial 2 & 61.9 & 51.6 & 18.0 & 57.3 & 46.2 & 18.9 & 60.3 & 42.8 & 18.1 \\
    & Trial 3 & 60.1 & 47.7 & 14.9 & 57.9 & 43.9 & 14.1 & 58.1 & 40.6 & 13.2 \\
    & Trial 4 & 58.6 & 49.5 & 12.4 & 53.3 & 40.5 & 15.2 & 55.2 & 41.5 & 13.5 \\
    & Trial 5 & 59.0 & 42.6 & 13.8 & 55.3 & 38.4 & 14.1 & 54.4 & 39.1 & 13.8 \\
    \midrule
    \multirow{6}{*}{Qwen-3-Reasoning} 
    & Trial 0 & 72.8 & 63.9 & 41.8 & 75.1 & 62.6 & 37.2 & 76.6 & 61.2 & 33.1 \\
    & Trial 1 & 58.8 & 51.1 & 18.5 & 60.9 & 47.2 & 15.3 & 60.6 & 42.1 & 15.6 \\
    & Trial 2 & 57.3 & 49.1 & 15.1 & 58.9 & 45.5 & 14.9 & 54.0 & 42.2 & 11.5 \\
    & Trial 3 & 56.3 & 45.7 & 14.7 & 54.6 & 42.1 & 12.9 & 58.9 & 37.4 & 13.8 \\
    & Trial 4 & 58.7 & 45.7 & 12.5 & 52.1 & 40.5 & 12.0 & 54.0 & 35.7 & 12.6 \\
    & Trial 5 & 55.7 & 44.5 & 12.2 & 53.9 & 37.1 & 11.6 & 52.5 & 35.6 & 10.7 \\
    \bottomrule
  \end{tabular}}
\end{table*}

\begin{table*}[t]
  \centering
  \caption{The factual accuracy of LLMs across multiple trials on 2-hop questions.}
  \resizebox{0.65\textwidth}{!}{
  \begin{tabular}{ll ccc ccc ccc}
    \toprule
    \multirow{2}{*}{LLM} & \multirow{2}{*}{Trial} & \multicolumn{3}{c}{Hum.} & \multicolumn{3}{c}{Soc. Sci.} & \multicolumn{3}{c}{STEM} \\
    \cmidrule(lr){3-5} \cmidrule(lr){6-8} \cmidrule(lr){9-11}
    & & Yes/No & MC & & Yes/No & MC & & Yes/No & MC  &\\
    \midrule
    \multirow{6}{*}{GPT-3.5} 
    & Trial 0 & 68.7 & 64.2 & & 66.2 & 50.3 & & 66.4 & 45.2 &\\
    & Trial 1 & 54.5 & 47.5 & & 54.8 & 37.7 & & 53.7 & 34.0 & \\
    & Trial 2 & 52.4 & 36.5 & & 54.0 & 35.1 & & 51.2 & 33.1 &\\
    & Trial 3 & 52.3 & 30.1 & & 53.0 & 32.5 & & 50.4 & 29.0 &\\
    & Trial 4 & 51.4 & 27.8 & & 51.9 & 30.6 & & 48.3 & 26.2 &\\
    & Trial 5 & 51.1 & 26.4 & & 51.0 & 29.4 & & 44.7 & 23.8 &\\
    \midrule
    \multirow{6}{*}{GPT-4o} 
    & Trial 0 & 74.4 & 74.8 & & 68.8 & 61.5 & & 70.5 & 49.6 &\\
    & Trial 1 & 62.6 & 52.1 & & 56.9 & 44.5 & & 58.7 & 42.3 &\\
    & Trial 2 & 60.4 & 45.9 & & 55.2 & 40.4 & & 56.0 & 35.5 &\\
    & Trial 3 & 58.5 & 41.5 & & 54.5 & 37.5 & & 54.4 & 32.5 &\\
    & Trial 4 & 57.5 & 38.0 & & 53.3 & 34.8 & & 52.3 & 30.1 &\\
    & Trial 5 & 57.1 & 35.6 & & 53.0 & 33.1 & & 51.7 & 28.3 &\\
    \midrule
    \multirow{6}{*}{GPT-4} 
    & Trial 0 & 73.4 & 72.4 & & 68.3 & 62.0 & & 69.2 & 50.2 & \\
    & Trial 1 & 62.2 & 55.7 & & 57.9 & 49.8 & & 54.8 & 43.7 &\\
    & Trial 2 & 58.4 & 47.2 & & 59.0 & 54.2 & & 53.7 & 37.3 &\\
    & Trial 3 & 56.4 & 41.5 & & 56.8 & 53.8 & & 53.0 & 34.7 &\\
    & Trial 4 & 56.4 & 38.2 & & 55.3 & 53.4 & & 51.9 & 32.1 &\\
    & Trial 5 & 55.8 & 35.9 & & 54.7 & 52.9 & & 51.1 & 30.3 &\\
    \midrule
    \multirow{6}{*}{DeepSeek-V3} 
    & Seed & 67.7 & 56.3 & & 66.2 & 45.2 & & 62.4 & 26.9 &\\
    & Trial 1 & 54.7 & 48.6 & & 53.9 & 38.4 & & 54.2 & 26.1 &\\
    & Trial 2 & 53.2 & 44.4 & & 51.9 & 35.3 & & 51.6 & 36.8 &\\
    & Trial 3 & 52.1 & 38.8 & & 51.1 & 32.4 & & 51.3 & 35.4 &\\
    & Trial 4 & 51.7 & 35.5 & & 50.5 & 30.8 & & 48.7 & 34.2 &\\
    & Trial 5 & 51.5 & 33.2 & & 50.0 & 28.7 & & 47.2 & 33.1 &\\
    \midrule
    \multirow{6}{*}{Claude-3.5-Haiku} 
    & Trial 0 & 70.5 & 55.7 & & 55.4 & 43.8 & & 68.1 & 36.4 &\\
    & Trial 1 & 59.5 & 38.8 & & 54.9 & 40.6 & & 57.7 & 37.3 &\\
    & Trial 2 & 56.7 & 38.3 & & 54.3 & 37.9 & & 54.1 & 32.3 &\\
    & Trial 3 & 56.0 & 34.5 & & 53.5 & 35.1 & & 53.7 & 30.0 &\\
    & Trial 4 & 54.7 & 32.8 & & 53.1 & 32.5 & & 52.0 & 28.9 &\\
    & Trial 5 & 53.6 & 31.8 & & 52.6 & 31.6 & & 50.2 & 28.0 &\\
    \midrule
    \multirow{6}{*}{Claude-Sonnet-4.0} 
    & Trial 0 & 70.2 & 68.5 & & 70.2 & 56.8 & & 66.1 & 40.2 &\\
    & Trial 1 & 55.7 & 48.2 & & 54.5 & 40.5 & & 52.0 & 39.9 &\\
    & Trial 2 & 54.4 & 36.6 & & 53.8 & 30.8 & & 50.2 & 34.8 &\\
    & Trial 3 & 54.4 & 31.0 & & 53.6 & 25.2 & & 49.1 & 31.0 &\\
    & Trial 4 & 53.5 & 26.8 & & 52.9 & 21.9 & & 47.9 & 28.2 &\\
    & Trial 5 & 52.6 & 24.0 & & 52.5 & 19.7 & & 47.7 & 26.2 &\\
    \midrule
    \multirow{6}{*}{Gemini-2.0} 
    & Trial 0 & 73.1 & 70.1 & & 67.6 & 59.4 & & 68.5 & 46.5 &\\
    & Trial 1 & 60.3 & 50.6 & & 52.8 & 42.9 & & 52.4 & 41.8 &\\
    & Trial 2 & 58.4 & 45.3 & & 51.6 & 37.3 & & 50.9 & 37.4 &\\
    & Trial 3 & 56.5 & 40.4 & & 51.7 & 34.7 & & 50.5 & 33.8 &\\
    & Trial 4 & 55.9 & 37.4 & & 50.7 & 32.3 & & 49.7 & 30.3 &\\
    & Trial 5 & 55.3 & 35.1 & & 50.5 & 30.5 & & 48.9 & 27.8 &\\
    \midrule
    \multirow{6}{*}{Qwen-3} 
    & Trial 0 & 69.3 & 63.4 & & 65.5 & 51.6 & & 65.0 & 42.5 &\\
    & Trial 1 & 59.3 & 49.3 & & 54.5 & 43.8 & & 53.5 & 39.8 &\\
    & Trial 2 & 55.6 & 36.8 & & 52.7 & 36.0 & & 43.9 & 23.6 &\\
    & Trial 3 & 54.8 & 31.0 & & 53.2 & 31.1 & & 33.2 & 23.5 &\\
    & Trial 4 & 54.0 & 30.6 & & 50.0 & 28.7 & & 24.2 & 18.5 &\\
    & Trial 5 & 53.7 & 30.5 & & 47.3 & 26.3 & & 42.8 & 16.9 &\\
    \midrule
    \multirow{6}{*}{Qwen-3-Reasoning} 
    & Trial 0 & 67.6 & 59.9 & & 65.8 & 44.0 & & 61.2 & 41.3 &\\
    & Trial 1 & 56.8 & 44.5 & & 53.3 & 39.5 & & 55.5 & 42.5 &\\
    & Trial 2 & 57.1 & 35.7 & & 52.2 & 29.5 & & 49.4 & 26.2 &\\
    & Trial 3 & 54.5 & 31.3 & & 51.8 & 29.4 & & 52.2 & 25.9 &\\
    & Trial 4 & 52.9 & 31.5 & & 51.2 & 25.9 & & 45.5 & 20.2 &\\
    & Trial 5 & 54.9 & 25.6 & & 51.0 & 25.5 & & 43.9 & 18.3 &\\
    \bottomrule
  \end{tabular}}
  \label{tab:llm-accuracy-trials-2hop}
\end{table*}

\section{Full Text of Human Annotation Instruction}
\subsection{Instructions for Question-Triplet Alignment (Section~\ref{humanano})}
Thank you for participating in this evaluation. Your task is to assess 200 questions generated from fact triplets to determine if they accurately reflect the triplet’s semantic meaning. For each assigned question-triplet pair, review the triplet and question, then record a Yes (question aligns with the triplet’s content and intent) or No (question misrepresents the triplet or deviates from instructions) in the provided options. Noted that all data will only be used as research purpose.

\subsection{Instructions for Factual Error Validation (Section~\ref{humanano2})}
Thank you for participating in this evaluation. Your task is to validate 100 failure cases identified by \methodname, where LLMs provided incorrect responses. For each question, triplet, and LLM response: (1) Answer using reliable internet sources. (2) Compare LLM response to the correct answer. (3) Discuss with annotators to resolve disagreements. (4) Annotate as valid error (incorrect LLM response) or false negative (correct LLM response). Data is used exclusively for research.

\section{Ethical Statement}
All participants were informed beforehand that their feedback would be used for research purposes and potentially published in this paper. No personal or private information was collected, and all data collected was anonymized to ensure privacy. Participation was voluntary, and the recruited volunteers were proficient in using LLMs to ensure informed contributions. 

\section{Beyond Evaluation — Using \methodname to Improve LLM Accuracy}
To complement our main study, we conducted a preliminary fine-tuning experiment using errors identified by \methodname. While not the focus of this work, the results indicate that these errors can be leveraged to improve factual accuracy in LLMs.

For fine-tuning the research models, we gathered 900 questions that were answered incorrectly by LLaMA-2-13B-Chat model and adopted a cross-format fine-tunning design. For example, if the model incorrectly answered an MC question about (Napoleon, native language, Corsican) during fine-tuning, we tested it with a Yes/No question on the same triplet. For the training set used for fine-tuning, we augmented it with the LIMA \cite{zhou2023lima} instruction tuning dataset so the model is able to answer questions in the expected format. We trained the LLaMA-2-13B-Chat model on 8 V100-32G GPUs using DeepSpeed Zero3, with a per-GPU batch size of 4, a learning rate of 2e-5, and a cosine learning rate schedule over 1 epoch. Post-fine-tuning, we re-evaluated the model with \methodname’s different format set.

The cross-format testing design provides evidence of factual learning rather than memorization of specific question-answer test cases. If the model merely memorized question-answer patterns, it would fail when tested with different question types on the same triplets. The consistent ~30\% improvement across different formats shows the model internalized the underlying knowledge, which is the expected and intended outcome of knowledge injection through fine-tuning on identified error cases. The result demonstrates that the factual errors identified by \methodname can substantially enhance the factual accuracy, resulting in a notable enhancement of 33.2\% for this fine-tuning method, from 35.3\% to 68.5\%. 

To examine whether this improvement comes at the expense of unrelated retained knowledge, we additionally evaluated the pre-SFT and post-SFT models on a held-out set of 500 randomly sampled Yes-No questions that were disjoint from the 900 questions used for fine-tuning. The two stages achieved nearly identical accuracies, 35.2\% before fine-tuning and 35.5\% after fine-tuning. This result suggests that, in our preliminary setting, the SFT procedure improved targeted factual errors without causing an observable degradation on unrelated knowledge of the same evaluation type.

We note, however, that this experiment is limited in scope: it only probes unrelated factual knowledge on a held-out Yes-No set and does not constitute a comprehensive evaluation of the model's broader general capabilities. A more complete study of trade-offs---including instruction following, reasoning, and other downstream abilities---is an important direction for future work.

\begin{table*}[h]
  \centering
  \caption{Coverage scores of LLMs across multiple trials on single-hop questions.}
  \label{tab:llm-coveragescore-trials}
  \resizebox{0.8\textwidth}{!}{
  \begin{tabular}{ll ccc ccc ccc}
    \toprule
    \multirow{2}{*}{LLM} & \multirow{2}{*}{Trial} & \multicolumn{3}{c}{Hum.} & \multicolumn{3}{c}{Soc. Sci.} & \multicolumn{3}{c}{STEM} \\
    \cmidrule(lr){3-5} \cmidrule(lr){6-8} \cmidrule(lr){9-11}
    & & Yes/No & MC & WH & Yes/No & MC & WH & Yes/No & MC & WH \\
    \midrule
    \multirow{6}{*}{GPT-3.5} 
    & Trial 0 & 0.357 & 0.360 & 0.359 & 0.395 & 0.391 & 0.400 & 0.312 & 0.306 & 0.254 \\
    & Trial 1 & 0.358 & 0.374 & 0.351 & 0.416 & 0.404 & 0.386 & 0.306 & 0.319 & 0.304 \\
    & Trial 2 & 0.406 & 0.431 & 0.392 & 0.461 & 0.434 & 0.416 & 0.357 & 0.379 & 0.345 \\
    & Trial 3 & 0.438 & 0.467 & 0.426 & 0.487 & 0.461 & 0.441 & 0.389 & 0.408 & 0.375 \\
    & Trial 4 & 0.464 & 0.489 & 0.449 & 0.518 & 0.493 & 0.466 & 0.417 & 0.429 & 0.397 \\
    & Trial 5 & 0.495 & 0.512 & 0.473 & 0.541 & 0.524 & 0.486 & 0.439 & 0.447 & 0.414 \\
    \midrule
    \multirow{6}{*}{GPT-4o} 
    & Trial 0 & 0.357 & 0.360 & 0.359 & 0.395 & 0.391 & 0.412 & 0.312 & 0.306 & 0.254 \\
    & Trial 1 & 0.379 & 0.385 & 0.368 & 0.431 & 0.407 & 0.421 & 0.325 & 0.307 & 0.327 \\
    & Trial 2 & 0.432 & 0.435 & 0.418 & 0.497 & 0.463 & 0.473 & 0.371 & 0.354 & 0.383 \\
    & Trial 3 & 0.471 & 0.474 & 0.446 & 0.525 & 0.506 & 0.500 & 0.412 & 0.401 & 0.413 \\
    & Trial 4 & 0.500 & 0.498 & 0.466 & 0.546 & 0.529 & 0.531 & 0.453 & 0.430 & 0.438 \\
    & Trial 5 & 0.521 & 0.521 & 0.488 & 0.566 & 0.545 & 0.550 & 0.472 & 0.457 & 0.458 \\
    \midrule
    \multirow{6}{*}{GPT-4} 
    & Trial 0 & 0.356 & 0.360 & 0.359 & 0.395 & 0.391 & 0.412 & 0.312 & 0.306 & 0.254 \\
    & Trial 1 & 0.369 & 0.378 & 0.365 & 0.450 & 0.436 & 0.397 & 0.327 & 0.354 & 0.338 \\
    & Trial 2 & 0.422 & 0.414 & 0.413 & 0.489 & 0.473 & 0.424 & 0.382 & 0.385 & 0.393 \\
    & Trial 3 & 0.448 & 0.439 & 0.441 & 0.513 & 0.506 & 0.445 & 0.413 & 0.404 & 0.421 \\
    & Trial 4 & 0.474 & 0.453 & 0.458 & 0.531 & 0.521 & 0.462 & 0.431 & 0.424 & 0.444 \\
    & Trial 5 & 0.495 & 0.470 & 0.479 & 0.542 & 0.535 & 0.480 & 0.449 & 0.439 & 0.459 \\
    \midrule
    \multirow{6}{*}{DeepSeek-V3} 
    & Trial 0 & 0.356 & 0.360 & 0.359 & 0.395 & 0.391 & 0.412 & 0.312 & 0.306 & 0.254 \\
    & Trial 1 & 0.374 & 0.370 & 0.360 & 0.426 & 0.415 & 0.379 & 0.277 & 0.363 & 0.332 \\
    & Trial 2 & 0.417 & 0.409 & 0.401 & 0.458 & 0.455 & 0.405 & 0.343 & 0.396 & 0.383 \\
    & Trial 3 & 0.454 & 0.433 & 0.436 & 0.483 & 0.487 & 0.423 & 0.400 & 0.415 & 0.413 \\
    & Trial 4 & 0.480 & 0.453 & 0.460 & 0.502 & 0.505 & 0.442 & 0.428 & 0.434 & 0.433 \\
    & Trial 5 & 0.504 & 0.472 & 0.482 & 0.520 & 0.520 & 0.457 & 0.453 & 0.449 & 0.448 \\
    \midrule
    \multirow{6}{*}{Claude-3.5-Haiku} 
    & Trial 0 & 0.356 & 0.360 & 0.359 & 0.395 & 0.391 & 0.412 & 0.312 & 0.306 & 0.254 \\
    & Trial 1 & 0.366 & 0.358 & 0.355 & 0.413 & 0.441 & 0.387 & 0.327 & 0.334 & 0.339 \\
    & Trial 2 & 0.403 & 0.394 & 0.395 & 0.449 & 0.479 & 0.423 & 0.377 & 0.367 & 0.379 \\
    & Trial 3 & 0.435 & 0.415 & 0.420 & 0.470 & 0.504 & 0.454 & 0.412 & 0.393 & 0.405 \\
    & Trial 4 & 0.447 & 0.435 & 0.436 & 0.487 & 0.521 & 0.472 & 0.429 & 0.415 & 0.427 \\
    & Trial 5 & 0.458 & 0.451 & 0.456 & 0.503 & 0.533 & 0.488 & 0.451 & 0.430 & 0.444 \\
    \midrule
    \multirow{6}{*}{Claude-Sonnet-4.0} 
    & Trial 0 & 0.356 & 0.360 & 0.359 & 0.395 & 0.391 & 0.412 & 0.312 & 0.306 & 0.254 \\
    & Trial 1 & 0.366 & 0.355 & 0.352 & 0.400 & 0.418 & 0.400 & 0.317 & 0.323 & 0.335 \\
    & Trial 2 & 0.423 & 0.412 & 0.396 & 0.443 & 0.463 & 0.430 & 0.357 & 0.367 & 0.370 \\
    & Trial 3 & 0.451 & 0.445 & 0.430 & 0.469 & 0.490 & 0.452 & 0.384 & 0.389 & 0.396 \\
    & Trial 4 & 0.474 & 0.469 & 0.454 & 0.485 & 0.511 & 0.476 & 0.404 & 0.406 & 0.416 \\
    & Trial 5 & 0.492 & 0.487 & 0.474 & 0.498 & 0.528 & 0.490 & 0.422 & 0.417 & 0.434 \\
    \midrule
    \multirow{6}{*}{Gemini-2.0} 
    & Trial 0 & 0.356 & 0.360 & 0.359 & 0.395 & 0.391 & 0.412 & 0.312 & 0.306 & 0.254 \\
    & Trial 1 & 0.382 & 0.369 & 0.355 & 0.443 & 0.414 & 0.405 & 0.345 & 0.338 & 0.302 \\
    & Trial 2 & 0.422 & 0.413 & 0.393 & 0.486 & 0.461 & 0.435 & 0.381 & 0.380 & 0.349 \\
    & Trial 3 & 0.449 & 0.434 & 0.423 & 0.514 & 0.486 & 0.463 & 0.405 & 0.409 & 0.378 \\
    & Trial 4 & 0.467 & 0.452 & 0.446 & 0.531 & 0.512 & 0.483 & 0.425 & 0.435 & 0.397 \\
    & Trial 5 & 0.482 & 0.471 & 0.465 & 0.541 & 0.534 & 0.501 & 0.441 & 0.449 & 0.414 \\
    \midrule
    \multirow{6}{*}{Qwen-3} 
    & Trial 0 & 0.356 & 0.360 & 0.359 & 0.395 & 0.391 & 0.412 & 0.312 & 0.306 & 0.254 \\
    & Trial 1 & 0.362 & 0.367 & 0.360 & 0.445 & 0.402 & 0.381 & 0.312 & 0.346 & 0.290 \\
    & Trial 2 & 0.397 & 0.404 & 0.400 & 0.482 & 0.436 & 0.410 & 0.364 & 0.376 & 0.330 \\
    & Trial 3 & 0.428 & 0.440 & 0.427 & 0.503 & 0.460 & 0.429 & 0.394 & 0.400 & 0.362 \\
    & Trial 4 & 0.454 & 0.459 & 0.453 & 0.521 & 0.470 & 0.445 & 0.418 & 0.420 & 0.436 \\
    & Trial 5 & 0.473 & 0.475 & 0.474 & 0.537 & 0.478 & 0.455 & 0.437 & 0.441 & 0.457 \\
    \midrule
    \multirow{6}{*}{Qwen-3-Reasoning} 
    & Trial 0 & 0.356 & 0.360 & 0.359 & 0.395 & 0.391 & 0.412 & 0.312 & 0.306 & 0.254 \\
    & Trial 1 & 0.372 & 0.355 & 0.353 & 0.409 & 0.411 & 0.383 & 0.327 & 0.342 & 0.292 \\
    & Trial 2 & 0.412 & 0.395 & 0.387 & 0.457 & 0.446 & 0.411 & 0.367 & 0.388 & 0.333 \\
    & Trial 3 & 0.446 & 0.423 & 0.417 & 0.477 & 0.483 & 0.432 & 0.393 & 0.412 & 0.360 \\
    & Trial 4 & 0.478 & 0.445 & 0.442 & 0.500 & 0.502 & 0.449 & 0.415 & 0.434 & 0.381 \\
    & Trial 5 & 0.502 & 0.464 & 0.464 & 0.521 & 0.512 & 0.465 & 0.437 & 0.449 & 0.403 \\
    \bottomrule
  \end{tabular}}
\end{table*}

\end{document}